# Quantum Computing: A Taxonomy, Systematic Review and Future Directions

Sukhpal Singh Gill[1], Adarsh Kumar[2*], Harvinder Singh[3], Manmeet Singh[4, 5], Kamalpreet Kaur[6], Muhammad Usman[7] and Rajkumar Buyya[8]

[1]School of Electronic Engineering and Computer Science, Queen Mary University of London, UK
[2]Department of Systemics, School of Computer Science, University of Petroleum and Energy Studies, Dehradun, India
[3]Department of Virtualization, School of Computer Science, University of Petroleum and Energy Studies, Dehradun, India
[4]Jackson School of Geosciences, University of Texas at Austin, USA
[5]Centre for Climate Change Research, Indian Institute of Tropical Meteorology (IITM), Pune, India
[6]Seneca International Academy, Seneca, Toronto, Canada
[7]School of Computing and Information Systems, The University of Melbourne, Parkville, Victoria, Australia
[8]Cloud Computing and Distributed Systems (CLOUDS) Laboratory, School of Computing and Information Systems, The University of Melbourne, Australia

[*]Corresponding Author: Adarsh Kumar, Department of Systemics, School of Computer Science, University of Petroleum and Energy Studies, Dehradun, Uttarakhand, 248007, India; email: adarsh.kumar@ddn.upes.ac.in

**Abstract:** Quantum computing is an emerging paradigm with the potential to offer significant computational advantage over conventional classical computing by exploiting quantum-mechanical principles such as entanglement and superposition. It is anticipated that this computational advantage of quantum computing will help to solve many complex and computationally intractable problems in several application domains such as drug design, data science, clean energy, finance, industrial chemical development, secure communications, and quantum chemistry. In recent years, tremendous progress in both quantum hardware development and quantum software/algorithm have brought quantum computing much closer to reality. Indeed, the demonstration of quantum supremacy marks a significant milestone in the Noisy Intermediate Scale Quantum (NISQ) era – the next logical step being the quantum advantage whereby quantum computers solve a real-world problem much more efficiently than classical computing. As the quantum devices are expected to steadily scale up in the next few years, quantum decoherence and qubit interconnectivity are two of the major challenges to achieve quantum advantage in the NISQ era. Quantum computing is a highly topical and fast-moving field of research with significant ongoing progress in all facets. A systematic review of the existing literature on quantum computing will be invaluable to understand the state-of-the-art of this emerging field and identify open challenges for the quantum computing community to address in the coming years. This article presents a comprehensive review of quantum computing literature and proposes taxonomy of quantum computing. The proposed taxonomy is used to map various related studies to identify the research gaps. A detailed overview of quantum software tools and technologies, post-quantum cryptography and quantum computer hardware development captures the current state-of-the-art in the respective areas. The article identifies and highlights various open challenges and promising future directions for research and innovation in quantum computing.

## 1. INTRODUCTION

In his famous lecture in 1982, Richard Feynman envisioned a quantum machine working on the laws of quantum mechanics which can simulate quantum physics, and in many ways, this is considered one of the initial conceptions of quantum computing [1]. He postulated that nature is not classical and therefore to simulate natural phenomena, one would need a computing device which works on quantum mechanical principles. Indeed, quantum computers offer such possibilities, where computing can exploit quantum mechanical properties such as entanglement and superposition to offer tremendous computational capabilities necessary for simulations of complex quantum systems. The initial progress towards developing quantum computer hardware was relatively slow, because the proposed quantum mechanical properties are only observed at the very fundamental scale of nature (e.g., electron spins or photon polarization), which were very challenging to manipulate due to technological limitations. However, in recent years, the field of quantum computing has rapidly progressed and emerged as one of the highly topical areas of research. Quantum computing has the potential to offer computational capabilities which will surpass existing supercomputers, and this has sparked huge interest from both industry and academia to build a world's first quantum machine. Today, many big companies such as IBM, Google, Microsoft, and Intel, as well as many ambitious start-up companies such as Rigetti and IonQ are actively perusing the race to develop a first large scale universal quantum computer. In parallel to quantum hardware development, the area of quantum software and quantum algorithm development has also seen tremendous progress in the last few years.

It is well known that in conventional classical digital computing, the information is stored and processed as bits which can take a definite binary value ('0' or '1'). The equivalent in quantum computing is known as quantum bit, or just qubit, which by the virtue of quantum mechanics could take values of '0', '1', or any superpositions of '0' and '1' (effectively being in both 0 and 1 states at once!). Quantum computers, therefore, can access an exponentially large Hilbert space (or computational space), where 'n' qubits could be in a superposition state of $2^n$ possible outcomes at any given time. This will allow quantum computers to tackle large scale space problems.

Developing a large-scale quantum computer has its own challenges. One of the major challenges in quantum hardware development arises from decoherence of qubits, whereby qubits lose their coherent properties via interaction with an environment. This implies that qubits in a superposition state will decohere to classical bits and therefore any quantum advantage will diminish. In "Noisy Intermediate Scale Quantum" (NISQ), 'noisy' mentions the fact that what is happening in the environment would disturb the devices. To exemplify, small changes in temperature, stray electric or magnetic fields can cause the quantum information in the computer to be degraded [2] [3]. Much of the ongoing research efforts in quantum computing are focused on overcoming errors in NISQ devices by developing efficient error correction protocols. A second major challenge is related to connectivity of qubits in today's quantum devices. It is related to relatively sparse connectivity of qubits in today's quantum devices as it becomes non-trivial to map large depth quantum circuits with many two-qubit gates which require inter-qubit interactions via direct couplings.

Despite technical challenges, NISQ quantum computers are already offering glimpses of computational capabilities. The recent demonstration of quantum supremacy by Google team is a significant milestone in the area of quantum computing [4]. An intense global race is now ongoing to achieve a first quantum computing application which solves a useful real-world problem that is intractable on classical computers – also known as 'quantum advantage'. To achieve this feat, a significant progress in both error-corrected quantum hardware and quantum algorithm development will be required in the coming years.

Quantum algorithms are being developed and benchmarked on NISQ devices at a rapid pace. In early 90's, there were only a few notable quantum algorithms such as Grover's and Shor's; however today hundreds of new quantum algorithms have been developed [5]. Among these, one of the most widely used class of quantum algorithms is Variation Quantum Algorithms such as Variational Quantum Eigensolver (VQE) [6] which are based on a combination of quantum and classical components. VQE algorithms have shown excellent results on NISQ devices for problems in quantum chemistry and quantum machine learning fields. A few other major categories of quantum algorithms include Algebraic (such as discrete log or verifying matrix products), search (such as Grover and amplitude amplification), and variational (such as quantum approximate optimization).

The full potential of quantum computing for real-world applications can only be realized when a large-scale fault-tolerant universal quantum computer will be available which requires several years of further development. However, the quantum speed-up on the existing NISQ era devices is already being accessed for prototype applications exhibiting promising results. Among these, variational quantum algorithms and quantum machine learning are two of the most active areas of research for NISQ devices. Quantum machine learning promises to speed up machine learning algorithms for analyzing the classical data. There have already been proposals for quantum principle component analysis, quantum support vector machine and quantum neural network. It is not yet fully established if quantum machine learning would offer superior computational efficiency when compared to classical machine learning implementations; however, recent work has shown promising results [7] [8]. Quantum computers consume less energy, therefore processing data intensive problems by quantum machine learning algorithms can reduce down energy cost, and the dependency on fossil-fuels will decrease [9].

For the implementation of quantum algorithms on NISQ devices, there are several well-known models such as the quantum circuit model for gate-based universal quantum computing, adiabatic quantum computing or quantum annealing and one-way quantum computer [2]. Among these, quantum circuit model is considered as the most practical pathway due to the possibility of re-programming quantum computers based on a target problem.

Quantum computing does not yet have its own high-level programming language. In the circuit model, the algorithms are processed by constructing quantum circuits which systematically apply available quantum gates or operations to find the desired solution.

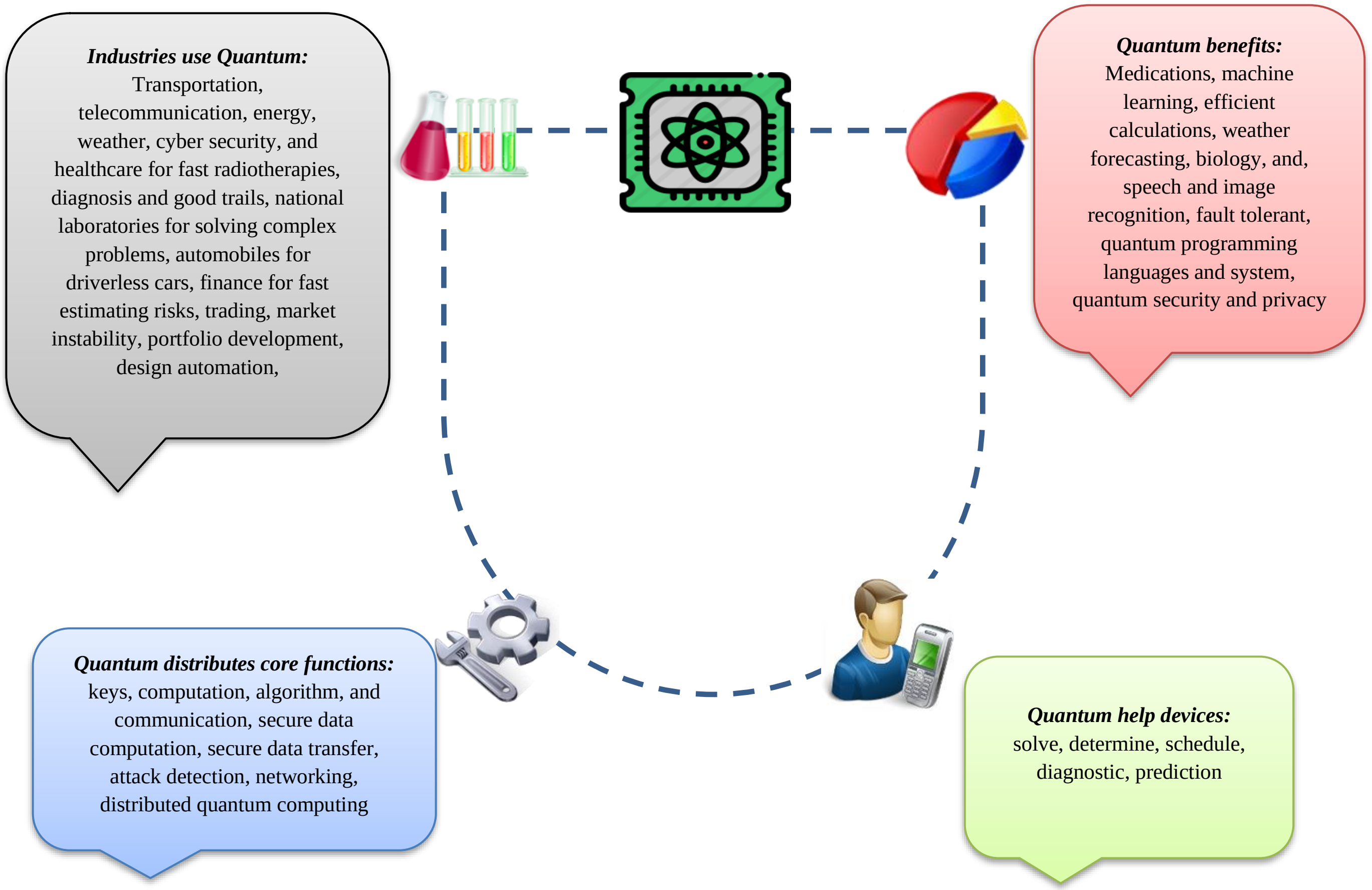

Figure 1: Quantum brings various advantages for the applications, application developers, and several industries by distributing the main functions

Another highly active area of research in the field of quantum computing is post-quantum secure communication. Cryptography is a technique which is used for hiding information from any unintended recipient [10]. Although quantum cryptography exploits quantum properties for sharing a quantum key (known as quantum key distribution or QKD) [11], post-quantum cryptography is still based on constructing classical cryptographic algorithms that hard to break by a quantum computer [12]. For post-quantum cryptography, major work is underway in developing many different techniques such as lattice-based, hash-based and code-based cryptography schemes [13].

Quantum computing is a rapidly progressing field of research with major developments happening all over the world towards many different aspects such as hardware development, software/algorithm development, error correction on NISQ devices, and applications. This article will provide a comprehensive and timely report on the recent progress and future directions, which will be beneficial for researchers as well as industry engineers working on a broad range of topics. As shown in Figure 1, quantum computing brings various advantages for the applications, application developers, and several industries by distributing the primary functions.

### 1.1 BASICS OF QUANTUM COMPUTING

The fundamental unit of classical computing is a bit, which can have two possible values '0' or '1' in binary format. Contrarily, in quantum computing the basic unit of information is a quantum bit or qubit. Qubits by the virtue of quantum mechanics can have a value of '0', '1' or both '0' and '1' simultaneously. Therefore, mathematically a qubit can be represented as $a|0\rangle + b|1\rangle$ where a and b are coefficients which allow mixing or superposition of '0' and '1' states. Figure 2 schematically shows the difference between a bit and a qubit in a superposition state.

The superposition of qubits provide access to a very large computational space which can solve many problems with large computational complexity. For example, a 3-bit number at any given time can have a single value from the set of eight possible values {000, 001, 010, 011, 100, 101, 110, 111}. However, a 3-qubit state can be placed in a superposition of all eight values: $a|000\rangle + b|001\rangle + c|010\rangle + d|011\rangle + e|100\rangle + f|101\rangle + g|110\rangle + h|111\rangle$. This implies that doubling the number of bits in a classical computing machine will only double the computational space, whereas the same can achieved by just adding one more qubit *i.e.* $2^3$ to $2^4$ by going from 3 to 4 qubits. This exponentially increasing computational space as a function of the number of qubits underpins the power of quantum computing which can handle very large dataset problems with only a small number of qubits. However, the loading of large data sets into quantum states is still an open question. Giovannetti et al. [304] presented the idea of using quantum random access memory, but its implementation on real quantum devices has not been demonstrated yet. Possible other solutions include using coreset constructions [305] and applying machine learning tools for the preparation of quantum states with learned data sets [306].

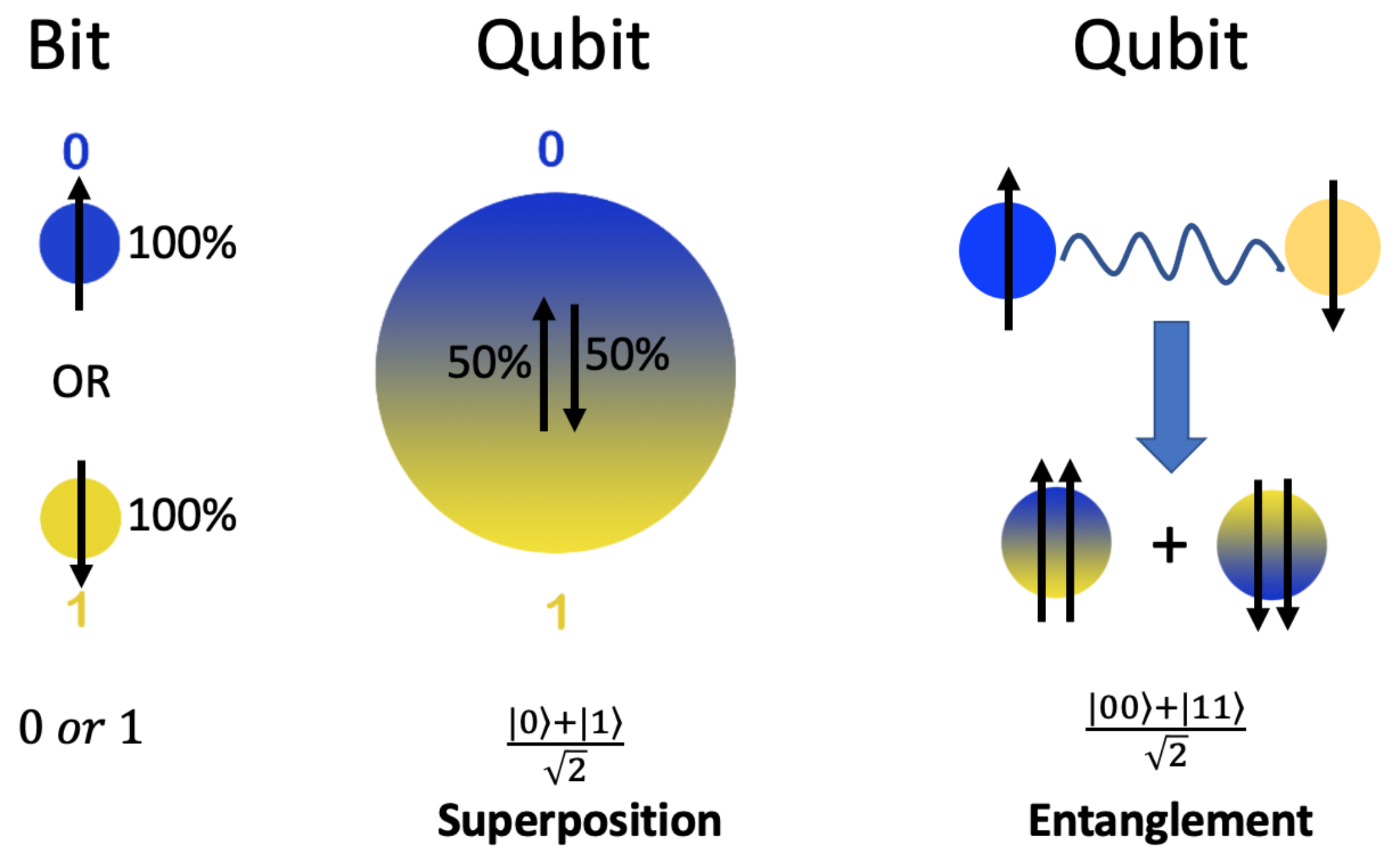

**Figure 2:** Illustration of a bit and qubit. (Left) A bit can take a value of '0' or '1' with 100% probability. (Middle) A qubit can be in a state of $|0\rangle$ or $|1\rangle$ or in a superposition state of both $|0\rangle$ and $|1\rangle$. Here, a qubit is illustrated in a superposition state, composed of 50% $|0\rangle$ and 50% $|1\rangle$. (Right) Illustration of two qubits in an entangled state. The properties of the two qubits in entangled state are linked to each other such that by looking (*i.e.* measuring) one of them, will reveal the other qubit, even when they are at physically large separations.

Another important property of quantum computing is entanglement which is illustrated in Figure 2. In contrast of classical bits where each bit value can be set independent of other bits, qubits can be placed in entangled states. In an entangled state, the properties of qubits are linked to each other, in spite of physical separation between them. Therefore, by measuring one qubit alter the properties of the other qubits which are in the same entangled state. Einstein famously called this 'spooky action at distance'. The entanglement is an important resource and can be exploited for dense coding and quantum simulation of correlated systems.

The simulation of a computational problem on a quantum computer typically follows a well-defined set of instructions. This includes the preparation of a superposition state which assigns equal probabilities to all possible outcomes. The implementation of quantum operations exploits superposition and entanglement properties in such a way that the probability of desired outcomes increases whereas the probabilities of other outcomes decrease. The last step in quantum computation is measurement, which leads to collapse of quantum state into the highest probability state providing the desired answer. The implementation of quantum algorithm makes sure that the desired outcome has probability very close to 1, with infinitesimally small probabilities for all other possibilities to achieve high fidelity outcomes.

### 1.2 MOTIVATION

The key motivation behind this comprehensive survey is to conduct a review of the existing literature on quantum computing. It covers the definition of quantum computing, its background, taxonomy, comparison of related studies based on taxonomy, quantum software tools and technologies, post-quantum cryptography and scalable quantum computer hardware. There is a need to identify open challenges and future research directions within the field of quantum computing.

### 1.3 RELATED SURVEYS AND OUR CONTRIBUTIONS

A few surveys were conducted on quantum computing in the literature. Savchuk and Fesenko [14] presented a general overview of quantum computing research, while Gyongyosi and Imre [15] discussed the fundamentals of quantum mechanics, such as quantum entanglement and quantum superposition. Dagmar et al. [16] presented a survey on quantum cryptography until the year 2007 but a lot of advanced research has been carried out in this field after that survey. Nour et al. [17] discussed advances in the quantum-theoretical approach to image processing applications only. An introduction to quantum computing for non-physicists was presented by Eleanor et al. [18]. Hamid et al. [19] proposed a survey on post-quantum lattice-based cryptography implementations, which is just one type of post-quantum cryptography. There is a need for a fresh systematic review which discusses everything from the definition of quantum computing to open challenges. Therefore, this study offers a systematic review of quantum computing literature, its taxonomy and maps the related studies based on this taxonomy. Further, detailed discussion on quantum software tools and technologies, post cryptography and industrial quantum computers is presented along with possible future directions. Table 1 shows the comparison of our survey with the existing surveys.

Table 1: Comparison of our survey with existing surveys

| Survey | General Overview | Basics of Quantum Computing | State of the art techniques | Taxonomy | Quantum Cryptography | Mapping of the Taxonomy | Quantum Software Tools and Technologies | Post-Quantum Cryptography | Scalable Quantum Computer Hardware | Future Research Directions |
|---|---|---|---|---|---|---|---|---|---|---|
| 1 | ✔(*) | ✔(*) | | | | | | | | |
| 2 | ✔(*) | ✔(*) | ✔ | | | | | | | |
| 3 | | | | | ✔(+) | | | | | |
| 4 | ✔(+) | ✔(+) | ✔(+) | | | | | | | |
| 5 | ✔(*) | | | | | | | | | |
| 6 | | | | | | | | ✔(+) | | |
| 7 | ✔(+) | ✔(+) | ✔(*) | ✔(*) | ✔(*) | ✔(*) | ✔(*) | ✔(*) | ✔(*) | ✔(*) |

1: Savchuk and Fesenko [14], 2: Gyongyosi and Imre [15], 3: Dagmar et al. [16], 4: Nour et al. [17], 5: Eleanor et al. [18], 6: Hamid et al. [19] and 7: Our Survey (This Paper). Note: (*: Means Comprehensive Discussion, +: Means Just an Overview)

## 1.4 ARTICLE STRUCTURE

The rest of the paper is organized as illustrated in Figure 3. Section 2 presents the building blocks and state of the art techniques for quantum computing. The taxonomy of quantum computing and its mapping is proposed in Section 3. Section 4 presents quantum software tools and technologies. The quantum and post-quantum cryptography are presented in Section 5. Section 6 presents the scalable quantum computer hardware. Section 7 highlights the future research directions. Section 8 concludes the paper. *Appendix A* notes the list of abbreviations.

### 2. BUILDING BLOCKS

Quantum mechanics concepts such as quantum interference, no-cloning theorem, quantum entanglement and quantum superposition are the underpinning principles of Quantum Computing (QC). In this section, we review the most recent literature on the technologies related to QC. The QC technologies are anticipated to offer significant speed-up in solving computational problems which otherwise are challenging when traditional computing techniques are used. In terms of the size of physical quantum devices, the quantum technologies are still in the phase of incubation.

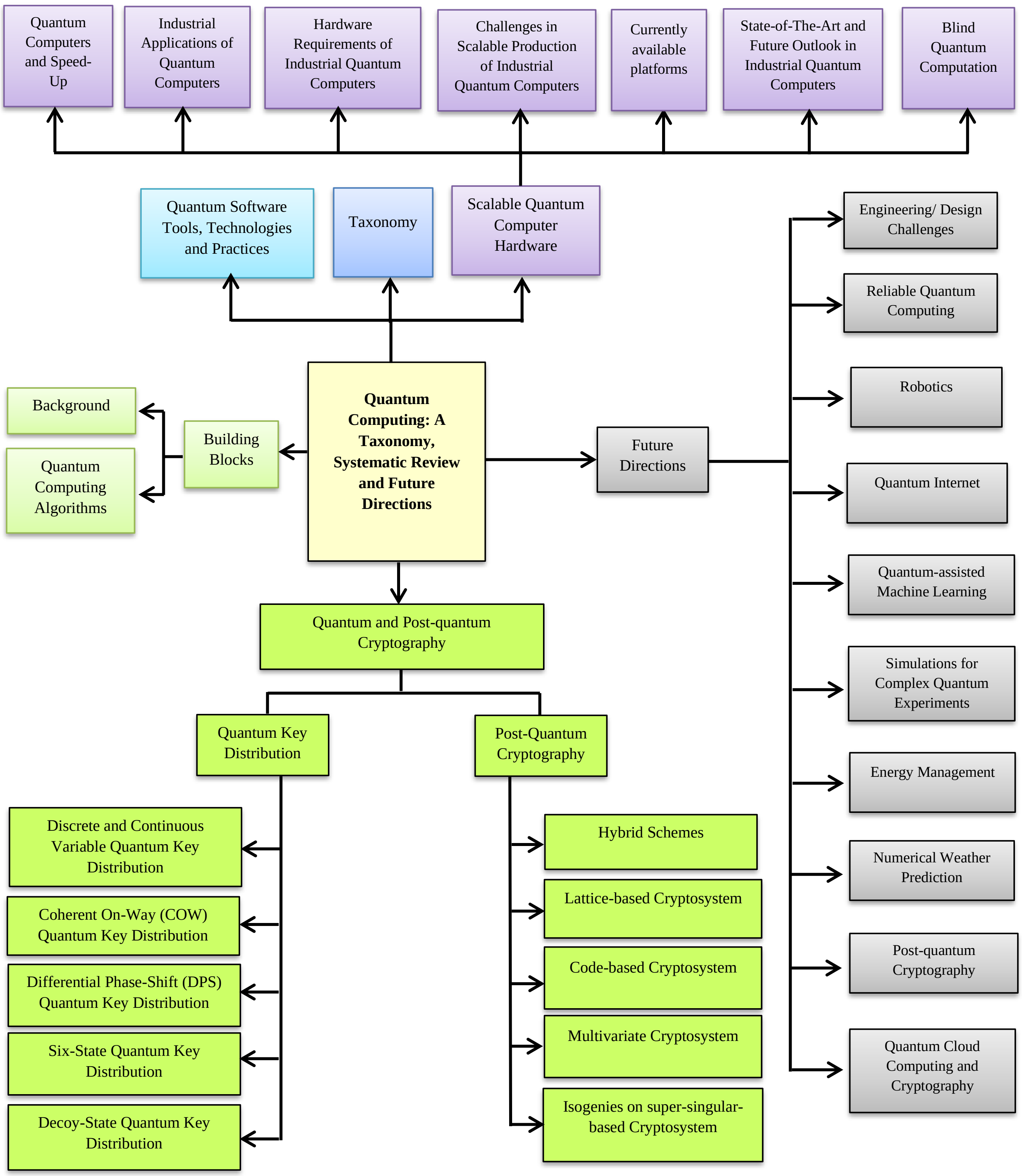

Figure 3: The Organization of this Survey

## 2.1 BACKGROUND

The basic building blocks of a large-scale quantum computer, as shown in Figure 4, consist of a quantum central processing unit, quantum gates, quantum control and measurement circuitry, quantum error detection and correction tools, and quantum memories.

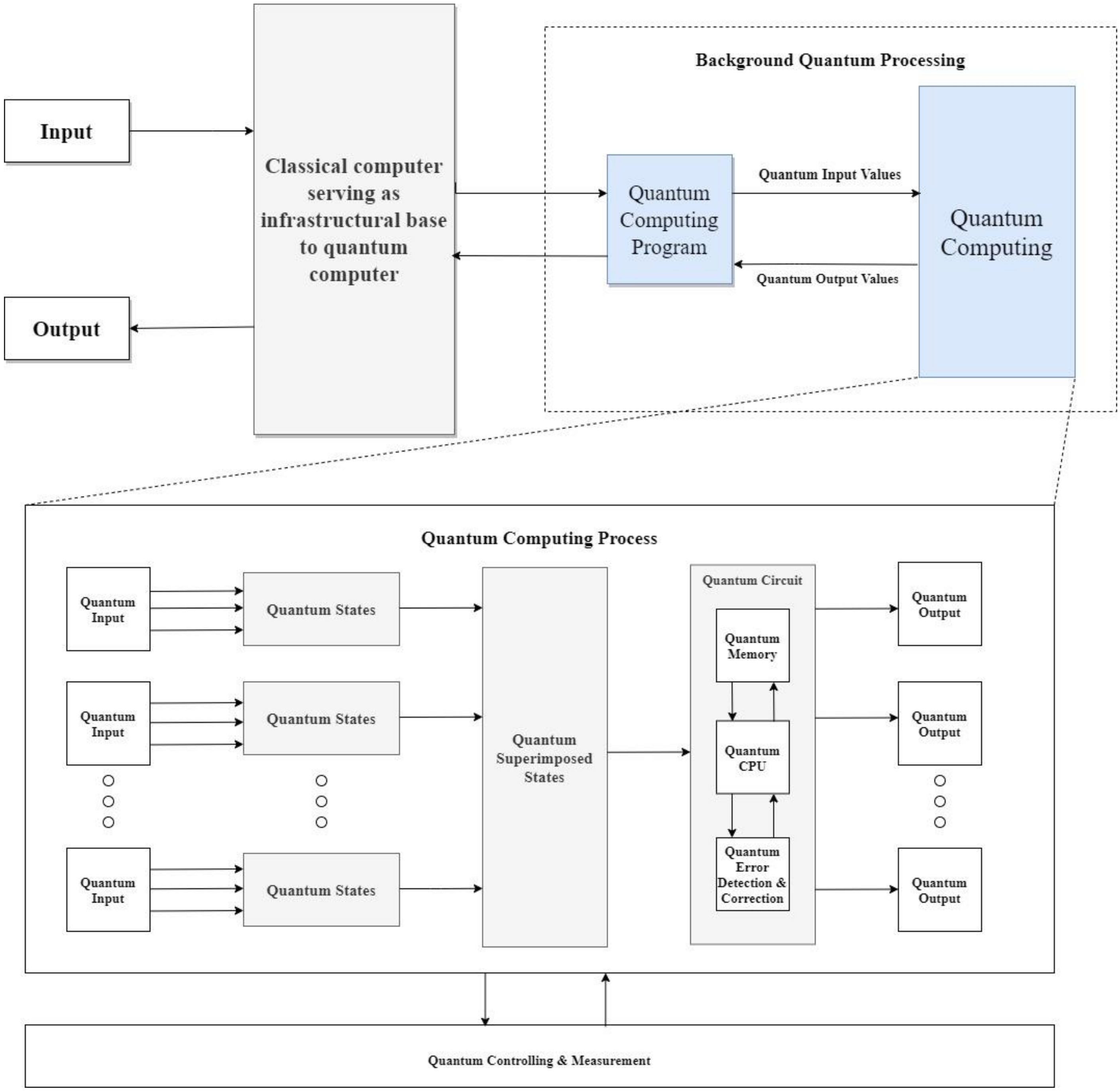

Figure 4: Basic building blocks of quantum computers

1. Quantum gates: The function of quantum gates in a quantum computer is to perform the operations that are unitary in nature [20]. Quantum gate is a combination of the multiple quantum circuits which uses quantum bits for their operations. Quantum logic gates are reversible in nature. Some examples of quantum gates are identity gate, pauli gates, phase shifter gates, 'hadamard gates, controlled gates and uncontrolled gates, rotation operator gates, swap gate and toffoli gate. All these gates differ in terms of a) how they are represented, b) number of qubits they are operating on. Quantum gates can be deployed in various arrangements such as shallow circuits [21], and instantaneous quantum polynomial-time circuit [22] depending upon the applications.

2. Quantum memory: The collection of multiple quantum states in various superposition arrangements constitutes quantum memories. Quantum memories use quantum registers to save the quantum states of a quantum circuit. Further the quantum states holds the important computational information known as qubits and qutrits. In the recent past, quantum memories have been realized using arrays of quantum states to form a stable quantum system [23].

3. Quantum processing unit: The Quantum Processing Unit (QPU) is an integral part of the quantum computer which works on the quantum computing principals to accomplish the task. These principles are based on the quantum mechanics, thus there is a significant difference between the conventional central processing unit and the QPU in terms of features. QPU stores the state of computation in terms of quantum mechanical state. It uses the quantum bus for communication amongst various other units of the quantum computer [24].

4. Quantum control and measurement circuitry: Quantum control and measurement mechanism is required in quantum computers for the proper monitoring of various manipulations of the quantum states and quantum computations while handling the error correction and detection processes [25].

5. Quantum error correction and detection tools: Quantum error detection and correction codes are used to locate and correct the errors that exist during the operations of the quantum gates. Quantum error correction is done to protect the quantum information from the errors that occurred because of quantum noise and decoherence. The error in quantum computers can be identified by using ancilla qubits without disturbing the information in data qubits. It is also important to note that the nature of errors detected in quantum computers is quite different when compared to traditional computing systems because the error can exist due to changes in amplitude or phase of a quantum state [26]. Quantum error correction and detection mechanism is required to achieve the fault tolerant quantum computation by not only dealing with the noise on stored quantum information, but also with faulty measurements, faulty quantum measurements and faulty quantum gates.

The core concept of quantum computing such as quantum logic gates, reversible computation idea of Fredkin and Bennet, quantum registers, qubits, Shor's factorization algorithm, quantum complexity and quantum entanglement has been discussed by Hey et al. [27][28]. In the same work, the experimental status has also been reviewed to get a better understanding of the quantum computer's physical implementation. To increase the computing performance of the classical computing system, various architectures of quantum computer that exist in the literature has been explored by Jain et al. [29]. The states of the quantum system are fragile in nature. When a quantum system interacts with its surrounding environment, the important quantum information about its states can leak. The leaked information cannot be recovered and used. The progressive deterioration of the state of a quantum system is known as system decoherence. With the existence of decoherence in the quantum system, the assessment of the quantum system will not produce the desired outcomes which can further results in failure of quantum algorithm. The architecture of quantum computer should be such that it should resolve this decoherence problem by proper management of errors that occur when performing quantum arithmetic computations. Kaiser et al. shares the lecture notes targeting those who are beginners to quantum computing field.

The basic idea was to provide the introduction of the fundamental concepts of quantum computing and explain how quantum topology enters the computation field. Buhrman et al. [30] performed a detailed survey of quantum computing techniques and explored its various applications in the distributed network framework. Gyongyosi and Imre [15] reviewed the most recent work done in the field of quantum computing. The experimental results of different quantum computing technologies have been demonstrated, and the problems related to it have been addressed. Savchuk et al. [14] emphasized on the concept of quantum computing which should be scalable. The existing quantum computer has been analyzed in detail for understanding its implementation. Further, it has been concluded that sufficient stress has not been laid by scientists and researchers on developing the scalable quantum computer. Zhang et al. [31] explored and revealed the concept of Quantum-inspired Evolutionary Algorithm (QEA) by merging two buzzwords, i.e. evolutionary algorithms and quantum computing. The basic architecture and system model of QEA has been explained and reviewed. The comparative analysis of various QEA has been discussed along with future research directions [31] [32]. Han et al. [33] proposed an algorithm that is evolutionary in nature and is inspired by the principles of Quantum-inspired Evolutionary Algorithm (QEA). The concept of the Quantum bits (Q-bit) and Quantum gates (Q-gate) have been applied enabling the algorithm to reach out to an optimal solution. For validation of QEA algorithm, its applicability for solving the knapsack problem is demonstrated, and the results have been compared with the traditional genetic algorithm. Rotteler et al. [34] provided an overview of quantum algorithms. They stated that quantum algorithms could be classified into three major categories namely amplitude amplification type algorithms, hidden subgroup type algorithms and the quantum algorithm that doesn't fall in the given two categories as the third category. All the three classifications have been studied to prove how quantum algorithms are different from traditional algorithms and how the computation speed will grow faster by using these quantum algorithms. Li et al. [35] discussed merging the elements of quantum mechanics with the intelligent nature-inspired algorithms to mark the new era of computing in the making. Quantum optimization and quantum learning are two classifications based on which the existing quantum algorithms were studied. Further, it was concluded that the nature-inspired quantum algorithms possess high potential when compared with classical-quantum computing algorithms.

For programming a quantum computer, special set of programming language tools are required. Sofge et al. [36] analyzed various programming language tools that are present in the market for quantum programmers. They carried out detailed comparative analysis among multiple tools available. Gay et al. [37] studied and reviewed the concept of quantum programming languages. Further, the design of quantum programming languages including their syntax, semantics and compilers for quantum computing have been discussed, and future research directions have been quoted. Menon et al. [38] pointed out the protocols required to provide the error-free translation of the abilities of the traditional computing system in contrast to the quantum computing system and vice versa. The existing simulators for quantum computing utilizing its capabilities to the fullest extent have been studied. Kumar et al. [39] discussed various components of quantum computing like qubits and quantum superposition. Quantum computers have been studied in terms of their efficiency and power. They picked two organizations from the list of organizations dealing with quantum computing and explored for their recent contributions in the field. Further, the research and development challenges related to quantum computing faced by these two organizations have been highlighted as future research challenges. Shaikh et al. [40] demonstrated the significance of big data analytics in quantum computing. Quantum machine learning algorithms that can scale up the processing speed of quantum processors by applying quantum walk in quantum Artificial Neural Networks (ANN) have been discussed. Yan et al. coined the concept of QIMP i.e. "Quantum Image Processing" which refers to the process of performing all kinds of manipulations on the quantum images for achieving multiple objectives. Further, different QIRs i.e. "Quantum Image Representation" which is the logical representation of the quantum images have been explored, and their applicability has been reviewed [41]. Roetteler et al. [42] stressed on the quantum security mechanisms and protocols involved in the various processes of a quantum computer. The comparative analysis of the cryptographic applications based on the quantum computing system and classical computing system has been done.

### 2.2 QUANTUM COMPUTING ALGORITHMS

Nobel laureate Richard Feynman was the first to postulate the idea of a quantum computer. The properties of quantum mechanics are leveraged by quantum computers and these properties form the basis of quantum computers. The quantum algorithms have come a long way starting from the simulations of quantum physics to a variety of applications in computer science. An industrial scale quantum computer will be a prized progress in achieving the processing power of its kind, which would have implications in various fields such as cybersecurity and others. The first quantum algorithm to find speed greater than that of a classical algorithm was proposed by Daniel Simon. Table 2 shows the comparison of quantum computing algorithms [28][43]-[53].

Table 2: Summary of Quantum Computing Algorithms

| Name | Year | Type | Objective |
|---|---|---|---|
| Deutsch–Jozsa Algorithm [296] | 1992 | Based on Quantum Fourier Transform | Problems requiring exponential queries |
| Bernstein–Vazirani Algorithm [297] | 1992 | | Efficient solutions of black-box problem |
| Simon's Algorithm [298] | 1994 | | Faster computation, speedup |
| Shor's Algorithm [43] | 1994 | | Integer factorization and discrete logarithm problems |
| Grover's Algorithm [44] | 1996 | Based on Amplitude Amplification | Searching unstructured database for marked entry |
| Quantum Counting [52] | 1998 | | Generalized search |
| Quantum Approximate Optimization Algorithm [53] | 2014 | Hybrid quantum/classical algorithm | Solution of graph theory problems |

## 3. TAXONOMY

In this section, quantum computing technologies are classified based on different types of features and operations. The various components of the quantum computing taxonomy are a) Basic characteristics, b) Algorithmic characteristics, c) Time and gate characteristics, and d) Other characteristics. A diagrammatic representation of the taxonomy of quantum computing is shown in Figure 5. We share a brief description of every element of quantum computing taxonomy.

- *Basic Characteristics:* The basic characteristics of quantum computing include elements like qubit implementation, classification based on quantum computing technology and performance metrics. The basic features of quantum computing are to explore how qubits can be implemented and represented. Qubit representation can be done either in stationary, flying or mobile ways. The stationary method is similar to traditional programming, whereas mobile approach resembles designing conventional circuits. Further, the ensemble computing and singleton computing is another classification based on the choice of quantum computing technology. An ensemble computing system is a group of quantum computers that are identical in specifications and performs the same set of functions. In contrast, the singleton computing system consists of a single quantum computer performing designated operations. The performance metrics forms the base of another classification of quantum computing techniques which has mechanical vibrations, fluorescence and concurrency as its attributes [38] [39].
- *Algorithmic Characteristics:* Quantum computing techniques can be realized by implementing quantum algorithms on the classical computing infrastructure. It is essential to discuss and categorize the quantum computing technologies on the basis of characteristics represented by quantum algorithms. The algorithmic elements of quantum computing technologies include: parallelism, aggregate count of qubits available, topologies, techniques for locating the qubits, and qubit operations.

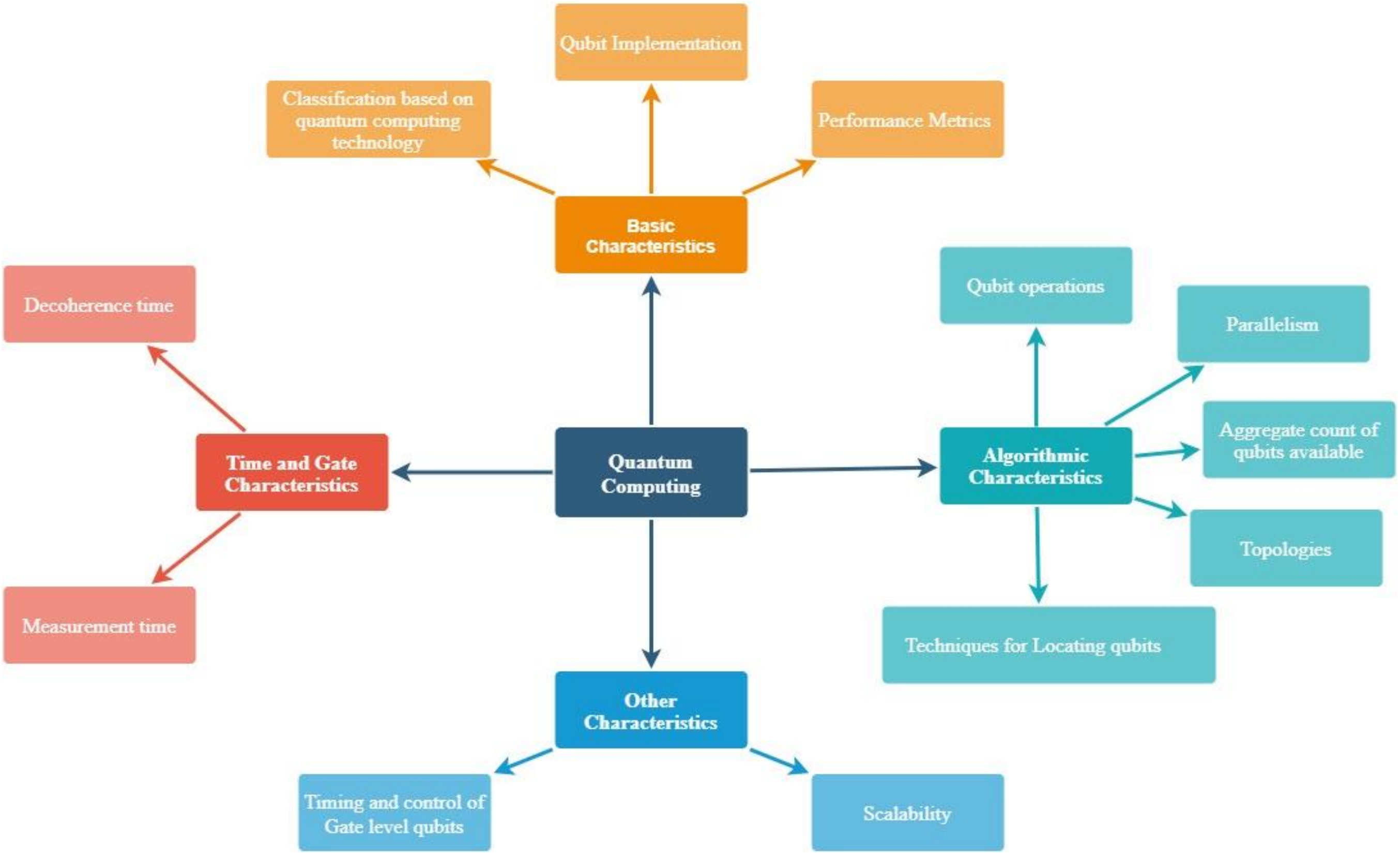

Figure 5: Taxonomy of quantum computing technology

Parallelism is the central feature because the parallel implementation of quantum gates is required to either prevent or minimize the qubits decoherence. The aggregate count of qubits available is another feature that helps in realizing the reliability and scalability of the quantum computer. The various possible arrangements of different physical devices in the architecture of the quantum computer is termed as its topologies. Architecture optimization is the primary concern as it enables the smooth flow of data and information among different physical units of the system. The addressing scheme for locating an individual qubit is logically very complex. This feature enables to explore the qubit states more specifically as far as the quantum computer physical implementation is concerned. Further, for performing any operation on qubits, they have to be moved from the address where they are stored to the location where the qubit gates are performing the action on them [34].

- *Time and Gate Characteristics:* The quantum computing technologies can be further classified on the basis of time and gate characteristics which include components such as decoherence time and measurement time. The decoherence time is given by the time until which a qubit can be kept in a specific state. The decoherence time is the topic of research in the field of quantum computing nowadays. Another essential characteristic of classification is the measurement time which is the time required to measure the qubit state precisely [7] [14].
- *Other Characteristics:* They include scalability, timing and control of gate-level qubits on which the classifications of quantum computing technologies have been done. All of the above-discussed features contribute towards the scaling of qubits to larger numbers. Meanwhile, it is recommended to use multiple qubits so that it will not always represent a single ion or photon. The changes in the qubit states are a continuous process with respect to time, hence computing an accurate timing of gates is critical. The arrival times of the qubits should be precisely adjusted while placing multiple qubits in their relative phases at the same time [15] [18].

The various quantum computing technologies are categorized based on our proposed taxonomy. The taxonomy-based mapping of quantum computing techniques is shown in Table 3. The quantum computing technologies considered for mapping are chosen based on the following criteria:

1. It represents the most recent and significant research work done in the field of quantum computing.
2. It should exhibit the fundamental characteristics defined in the taxonomy of quantum computing which forms the basis of mapping.

Table 3: Taxonomies-based mapping of quantum computing techniques

| Work | QI | CQCT | PM | P | ACQA | T | TLG | QO | DT | MT | TCGLQ | S |
|---|---|---|---|---|---|---|---|---|---|---|---|---|
| Weitenberg et al. [54] | ✓ | ✓ | ✓ | ✓ | ✓ | ✕ | ✓ | ✕ | ✓ | ✕ | ✓ | ✓ |
| Tomza et al. [55] | ✓ | ✓ | ✓ | ✓ | ✓ | ✓ | ✓ | ✕ | ✓ | ✕ | ✓ | ✓ |
| Gorman et al. [56] | ✓ | ✓ | ✓ | ✓ | ✓ | ✕ | ✓ | ✓ | ✓ | ✓ | ✓ | ✓ |
| Compagno et al. [57] | ✓ | ✓ | ✓ | ✕ | ✕ | ✕ | ✓ | ✕ | ✕ | ✕ | ✓ | ✓ |
| sSchaal et al. [58] | ✓ | ✓ | ✓ | ✕ | ✓ | ✕ | ✓ | ✕ | ✕ | ✕ | ✓ | ✓ |
| Zwanenburg et al. [59] | ✓ | ✓ | ✓ | ✕ | ✓ | ✕ | ✕ | ✓ | ✓ | ✕ | ✓ | ✓ |
| Veldhorst et al. [60] | ✓ | ✓ | ✓ | ✕ | ✓ | ✕ | ✕ | ✓ | ✕ | ✕ | ✓ | ✓ |
| Mizuta et al. [61] | ✓ | ✓ | ✓ | ✕ | ✓ | ✕ | ✕ | ✕ | ✕ | ✕ | ✕ | ✕ |
| Albornoz et al. [62] | ✓ | ✓ | ✓ | ✕ | ✓ | ✕ | ✕ | ✓ | ✕ | ✕ | ✓ | ✕ |

Abbreviations – QI: Qubit Implementation, CQCT: Classification based on quantum computing technology, PM: Performance Metrics, P: Parallelism, ACQA: Aggregate count of qubits available, T: Topologies, TLG: Techniques for Locating qubits, QO: Qubit operations, DT: Decoherence time, MT: Measurement time, TCGLQ: Timing and control of Gate level qubits, S: Scalability.

## 4. QUANTUM SOFTWARE TOOLS, TECHNOLOGIES AND PRACTICES

In comparison to the fields of quantum hardware development and quantum simulations, the area of quantum software development is relatively new and less established. Recently quantum software tools are being developed at a rapid pace with many quantum software packages now available from different platforms/sources such as Google, IBM, Microsoft, and D-Wave. These software tools are still at relatively low level such as at the level of assembly language; high-level quantum programming analogous to classical programming tools such as C++ and Java are not yet available.

Table 4 shows the comparative analysis of quantum tools available to-date. The analysis performed in table 4 presents the current scenario and it may change in future in highly dynamic quantum world and its growth. The parameters used for comparative analysis are briefly explained as follows: (i) a *library* is considered as a collection of functions or classes designed for quantum information and similar computations, (ii) a *tool* is a piece of software that can simulate quantum computing or associated calculations, (iii) the quantum computing libraries, tools, or techniques are found to be either *open-source*, *commercial* or *freeware*, (iv) Graphical User Interface *(GUI)-based* quantum tools are available that ease the job of circuit designing, programming and displaying the results for users, (v) many GUI-based tools can display the results either in *two or three-dimensions*, (vi) additionally, many tools have *command-line usage* where instructions are predefined to connect the gates, design the inputs and observe the outputs, (vii) *quantum gates* are analogous to conventional logic gates for quantum computers. Few examples of quantum gates include Hadamard, phase shifter, controlled, uncontrolled, and Controlled NOT (CNOT) gate. (viii) In this work, a review of most of the quantum programming tools available for academic or research work, used for *simulation* rather than *real-implementation*, is performed (ix) while exploring the quantum tools, it has been observed that many of such tools provide an existing implementation of quantum algorithms. Few of these algorithms include Shor, Deutsch-Jozsa, Simon, Quantum phase estimation, Hidden subgroup, and Grover algorithms [45] [50]. These algorithms are classified in one of the following categories: quantum Fourier transform, amplitude amplification, quantum walks, Bounded-error Quantum Polynomial time (BQP)-complete, and hybrid quantum/classical, (x) gates scheduling, and parallelism is vital for circuit designing that is analogous to quantum computer operations. This concept speed-up the operations in quantum computing, and (xi) most of the quantum gates require matrix operations for their computations. This matrix and associated operations are incorporated in many tools. Table 4 shows the comparative analysis of quantum tools with the above parameters. This comparative analysis shows the programming languages used in the tools as well.

Table 4: Comparative analysis of Software Tools and Technologies

| **Tool/Technique Name** | **A** | **B** | **C** | **D** | **E** | **F** | **G** | **H** | **I** | **J** | **K** | **L** | **M** | **N** | **O** | **Underpinning Programming Language** |
|---|---|---|---|---|---|---|---|---|---|---|---|---|---|---|---|---|
| QuEST [63] | ✔ | ✔ | ✔ | ✖ | ✖ | ✖ | ✖ | ✖ | ✔ | ✔ | ✔ | ✖ | ✖ | ✖ | ✖ | C |
| Staq [64] | ✔ | ✔ | ✔ | ✖ | ✖ | ✖ | ✖ | ✖ | ✔ | ✔ | ✔ | ✖ | ✖ | ✖ | ✖ | C++ |
| Scaffold/ScaffCC [65] | ✔ | ✔ | ✔ | ✖ | ✖ | ✖ | ✖ | ✖ | ✔ | ✖ | ✔ | ✖ | ✔ | ✖ | ✖ | Scaffold |
| Qrack [66] | ✔ | ✔ | ✔ | ✖ | ✖ | ✖ | ✖ | ✖ | ✔ | ✖ | ✔ | ✖ | ✖ | ✖ | ✖ | C++ |
| QX Simulator [50] | ✔ | ✔ | ✔ | ✖ | ✖ | ✖ | ✖ | ✖ | ✔ | ✖ | ✔ | ✖ | ✖ | ✔ | ✖ | Quantum Code |
| Quantum++ [67] | ✔ | ✔ | ✔ | ✖ | ✖ | ✖ | ✖ | ✖ | ✔ | ✖ | ✔ | ✖ | ✖ | ✖ | ✖ | C++ |
| QMDD [68] | ✔ | ✖ | ✔ | ✖ | ✖ | ✖ | ✖ | ✖ | ✔ | ✖ | ✔ | ✖ | ✖ | ✖ | ✔ | C++ |
| CHP [69] | ✔ | ✖ | ✔ | ✖ | ✖ | ✖ | ✖ | ✖ | ✔ | ✔ | ✔ | ✖ | ✔ | ✖ | ✖ | C |
| Eqcs [70] | ✔ | ✖ | ✔ | ✖ | ✖ | ✖ | ✖ | ✖ | ✔ | ✔ | ✔ | ✖ | ✖ | ✖ | ✖ | C |
| LanQ [71] | ✔ | ✖ | ✔ | ✖ | ✖ | ✖ | ✖ | ✖ | ✔ | ✔ | ✔ | ✖ | ✔ | ✔ | ✖ | LanQ |
| libquantum (C) [72] / (C++)[73] | ✔ | ✖ | ✔ | ✖ | ✖ | ✖ | ✖ | ✖ | ✔ | ✔ | ✔ | ✖ | ✔ | ✔ | ✖ | C, C++ |
| Open Qubit [74] | ✔ | ✖ | ✔ | ✖ | ✖ | ✖ | ✖ | ✖ | ✔ | ✔ | ✔ | ✖ | ✔ | ✖ | ✖ | C++ |
| Quantum Programming Studio [75] | ✖ | ✔ | ✔ | ✖ | ✖ | ✖ | ✖ | ✔ | ✖ | ✔ | ✔ | ✖ | ✔ | ✔ | ✔ | Javascript |
| Qubit Workbench [76] | ✖ | ✔ | ✖ | ✔ | ✖ | ✔ | ✖ | ✔ | ✖ | ✔ | ✔ | ✖ | ✔ | ✖ | ✔ | -- |
| Linear AI [77] | ✔ | ✖ | ✖ | ✖ | ✔ | ✔ | ✖ | ✖ | ✔ | ✔ | ✔ | ✖ | ✔ | ✖ | ✔ | Mathematica |
| QCAD [78] | ✖ | ✔ | ✖ | ✖ | ✔ | ✔ | ✔ | ✔ | ✖ | ✔ | ✔ | ✖ | ✖ | ✖ | ✔ | -- |
| qsims [79] | ✔ | ✔ | ✔ | ✔ | ✖ | ✖ | ✖ | ✖ | ✔ | ✔ | ✔ | ✖ | ✔ | ✔ | ✖ | C++ |
| Q-gol [80] | ✔ | ✔ | ✔ | ✖ | ✖ | ✔ | ✔ | ✔ | ✖ | ✔ | ✔ | ✖ | ✔ | ✖ | ✖ | CaML |
| QOCS [81] | ✔ | ✔ | ✔ | ✖ | ✖ | ✖ | ✖ | ✖ | ✔ | ✔ | ✔ | ✖ | ✔ | ✖ | ✖ | OCaML |
| Q++ [82] | ✔ | ✖ | ✔ | ✖ | ✖ | ✖ | ✖ | ✖ | ✔ | ✔ | ✔ | ✖ | ✔ | ✖ | ✔ | C++ |
| Qinf [83] | ✔ | ✔ | ✔ | ✖ | ✖ | ✖ | ✖ | ✖ | ✖ | ✔ | ✔ | ✖ | ✖ | ✖ | ✔ | Maxima |
| Quantum Fog [84] | ✔ | ✔ | ✔ | ✖ | ✖ | ✔ | ✖ | ✔ | ✖ | ✔ | ✔ | ✖ | ✔ | ✔ | ✔ | -- |
| SimQubit [85] | ✔ | ✔ | ✔ | ✖ | ✖ | ✔ | ✖ | ✔ | ✖ | ✔ | ✔ | ✖ | ✔ | ✖ | ✔ | C++ |
| Q-Kit [86] | ✖ | ✔ | ✖ | ✖ | ✔ | ✔ | ✖ | ✔ | ✖ | ✔ | ✔ | ✖ | ✔ | ✔ | ✔ | -- |
| Bloch Sphere [87] | ✔ | ✔ | ✔ | ✖ | ✖ | ✔ | ✔ | ✔ | ✖ | ✔ | ✔ | ✖ | ✔ | ✖ | ✔ | Java |
| BackupBrain [88] | ✔ | ✔ | ✔ | ✖ | ✖ | ✔ | ✖ | ✖ | ✖ | ✔ | ✔ | ✖ | ✔ | ✖ | ✔ | Javascript |
| Quantum Circuit [89] | ✔ | ✔ | ✔ | ✖ | ✖ | ✖ | ✖ | ✖ | ✔ | ✔ | ✔ | ✖ | ✔ | ✖ | ✔ | Javascript |
| Jsquis [90] | ✔ | ✔ | ✔ | ✖ | ✖ | ✖ | ✖ | ✖ | ✔ | ✔ | ✔ | ✖ | ✖ | ✖ | ✖ | Javascript |
| QSWalk.jl [91] | ✔ | ✔ | ✔ | ✖ | ✖ | ✔ | ✖ | ✖ | ✔ | ✔ | ✔ | ✖ | ✖ | ✖ | ✖ | Julia |
| QuantumOptics.jl [92] | ✔ | ✔ | ✔ | ✖ | ✖ | ✔ | ✔ | ✖ | ✔ | ✖ | ✔ | ✖ | ✔ | ✖ | ✔ | Julia |
| QuantumWalk,jl [93] | ✔ | ✔ | ✔ | ✖ | ✖ | ✔ | ✖ | ✖ | ✔ | ✔ | ✔ | ✖ | ✔ | ✖ | ✖ | Julia |
| Feynman [94] | ✔ | ✔ | ✔ | ✖ | ✖ | ✔ | ✖ | ✖ | ✔ | ✔ | ✔ | ✖ | ✔ | ✔ | ✖ | Maple |
| OpenQUACS [95] | ✔ | ✔ | ✔ | ✖ | ✖ | ✔ | ✖ | ✖ | ✔ | ✔ | ✔ | ✖ | ✔ | ✔ | ✖ | Maple |
| Quantavo [96] | ✔ | ✔ | ✔ | ✖ | ✖ | ✖ | ✖ | ✖ | ✔ | ✔ | ✔ | ✖ | ✔ | ✔ | ✖ | Maple |
| QDENSITY [97] | ✔ | ✔ | ✔ | ✖ | ✖ | ✔ | ✖ | ✖ | ✔ | ✔ | ✔ | ✖ | ✖ | ✖ | ✔ | Mathematica |
| Quantum [98] | ✔ | ✔ | ✖ | ✖ | ✔ | ✔ | ✖ | ✖ | ✔ | ✔ | ✖ | ✖ | ✔ | ✔ | ✔ | Mathematica |
| QuantumUtils [99] | ✔ | ✔ | ✔ | ✖ | ✖ | ✖ | ✖ | ✖ | ✔ | ✔ | ✔ | ✖ | ✖ | ✔ | ✔ | Mathematica |
| Qi [100] | ✔ | ✔ | ✔ | ✖ | ✖ | ✖ | ✖ | ✖ | ✔ | ✔ | ✔ | ✔ | ✖ | ✖ | ✖ | Mathematica |
| M-fun [101] | ✔ | ✖ | ✔ | ✖ | ✖ | ✔ | ✖ | ✖ | ✔ | ✔ | ✔ | ✖ | ✖ | ✖ | ✖ | MATLAB/Octave |
| Quantencomputer [102] | ✔ | ✔ | ✖ | ✖ | ✖ | ✖ | ✖ | ✖ | ✔ | ✔ | ✔ | ✖ | ✔ | ✖ | ✖ | MATLAB |
| Drqubit [103] | ✖ | ✖ | ✔ | ✖ | ✔ | ✖ | ✖ | ✖ | ✔ | ✔ | ✔ | ✖ | ✖ | ✔ | ✖ | MATLAB |
| Qubit4Matlab [104] | ✔ | ✖ | ✔ | ✖ | ✖ | ✖ | ✖ | ✖ | ✔ | ✔ | ✔ | ✖ | ✔ | ✔ | ✔ | MATLAB |
| QuIDE [105] | ✖ | ✔ | ✔ | ✖ | ✖ | ✔ | ✖ | ✔ | ✖ | ✔ | ✔ | ✖ | ✔ | ✔ | ✔ | .NET |
| Quantum.NET [106] | ✔ | ✔ | ✔ | ✖ | ✖ | ✖ | ✖ | ✖ | ✔ | ✔ | ✔ | ✖ | ✖ | ✖ | ✖ | .NET |
| Qubit Workbench [107] | ✖ | ✔ | ✖ | ✖ | ✔ | ✔ | ✖ | ✔ | ✖ | ✔ | ✔ | ✖ | ✖ | ✖ | ✖ | _ |
| Cirq [108] | ✔ | ✔ | ✔ | ✖ | ✖ | ✖ | ✖ | ✖ | ✔ | ✔ | ✔ | ✔ | ✔ | ✔ | ✔ | Python |
| ProjectQ [109] | ✔ | ✔ | ✔ | ✘ | ✘ | ✘ | ✘ | ✘ | ✔ | ✔ | ✔ | ✘ | ✔ | ✔ | ✔ | Python |
| QCircuits [110] | ✔ | ✔ | ✔ | ✘ | ✘ | ✘ | ✘ | ✘ | ✔ | ✔ | ✔ | ✘ | ✔ | ✔ | ✔ | Python |
| Qiskit [111] | ✔ | ✔ | ✔ | ✘ | ✘ | ✔ | ✘ | ✘ | ✔ | ✔ | ✔ | ✔ | ✔ | ✔ | ✔ | Python |
| OpenQasm [112] | ✔ | ✔ | ✔ | ✘ | ✘ | ✘ | ✘ | ✘ | ✔ | ✔ | ✔ | ✘ | ✘ | ✘ | ✔ | QASM |
| QCGPU [113] | ✔ | ✔ | ✔ | ✘ | ✘ | ✘ | ✘ | ✘ | ✔ | ✔ | ✔ | ✘ | ✘ | ✔ | ✘ | Rust & OpenCL |
| QIO [114] | ✔ | ✘ | ✔ | ✘ | ✘ | ✘ | ✘ | ✘ | ✔ | ✔ | ✔ | ✘ | ✘ | ✘ | ✔ | Qio + Haskell |
| Qchas [115] | ✔ | ✘ | ✔ | ✘ | ✘ | ✘ | ✘ | ✘ | ✔ | ✔ | ✔ | ✘ | ✘ | ✘ | ✔ | Haskell |
| Quantum User Interface [116] | ✘ | ✔ | ✔ | ✘ | ✘ | ✔ | ✘ | ✔ | ✘ | ✔ | ✔ | ✘ | ✘ | ✘ | ✘ | Protobuf |
| Quantum Development Kit (QDK) [117] | ✔ | ✔ | ✔ | ✘ | ✘ | ✔ | ✘ | ✘ | ✔ | ✔ | ✔ | ✘ | ✔ | ✔ | ✔ | Python, Q# |

**A: Library, B: Toolkit, C: Open Source, D: Commercial, E: Freeware, F: GUI-based, G: 3D visualization, H: Drag & Drop Support, I: Command-Line Usage J: Support for Quantum Gates, K: Simulation, L: Real-Implementation. M: Built-in quantum algorithm support, N: Gates scheduling & Parallelism, O: Diagram or Matrix support**

Figure 6 shows the major software engineering practices observed in recent studies of quantum computing world. These practices are briefly explained as follows [260] [261] [262].

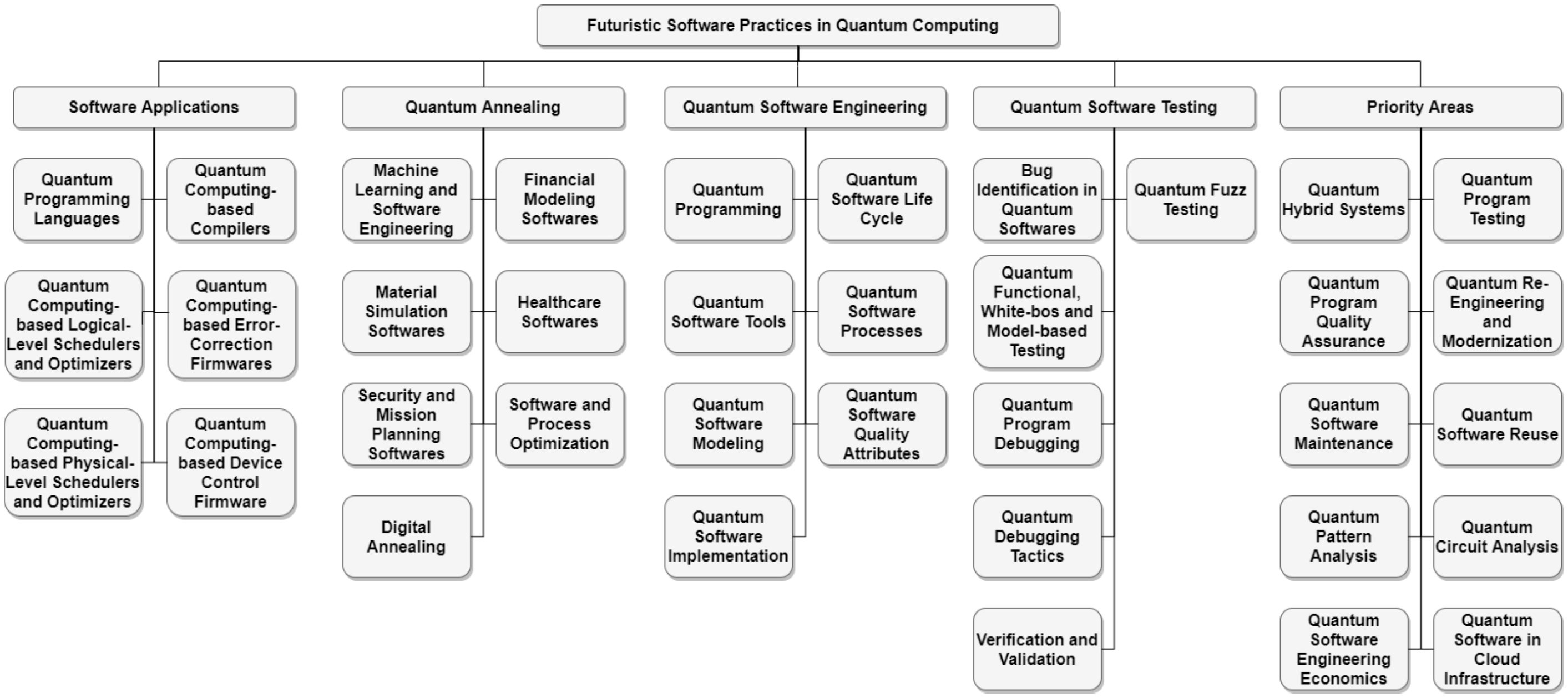

Figure 6: Futuristic Software Practices in Quantum Computing World

- ***Software Applications:*** In software applications, quantum programming languages, quantum computing-based compilers, quantum computing-based logical level schedulers and optimizers, quantum computing-based error-correction firmwares, quantum computing-based physical level schedulers and optimizers, and quantum computing-based device control firmware are major areas observed in recent studies. Figure 7 shows the classification of software applications in quantum computing.

- ***Quantum Programming Languages:*** Various studies are performed to discuss the importance and challenges of quantum programming languages [263]-[269]. According to Heim et al. [264] and Gay et al. [265], Important aspects to study in this domain includes (i) Programming Language Designs (PLD), (ii) Programming Language Semantics (PLS), (iii)Programming Language Compilation (PLC), (iv) Commuting Operations (CO), (v) Controlled Operations (COp), Adjoint Operations (AO), and Clean and borrowed Qubits (CbQ) [266] [267] [268] 269]. The important challenges in quantum programming languages include [263]-[269]: (i) to program infinite data types in programming languages that ensure storing of infinite quantum-data, (iv) to design quantum concurrency systems that support potential applications, (v) to provide quantum computation supported virtual machines, (vi) to design algorithms that support customized error correction techniques in programming languages, (vii) to focus on programming language designs like imperative, functional, and other languages and λ-calculi, (vii) to apply linear logics in programming languages-based applications, (viii) to apply domain-theoretic semantic techniques in programming languages, and (ix) to explore new semantic techniques in futuristic quantum programming languages.

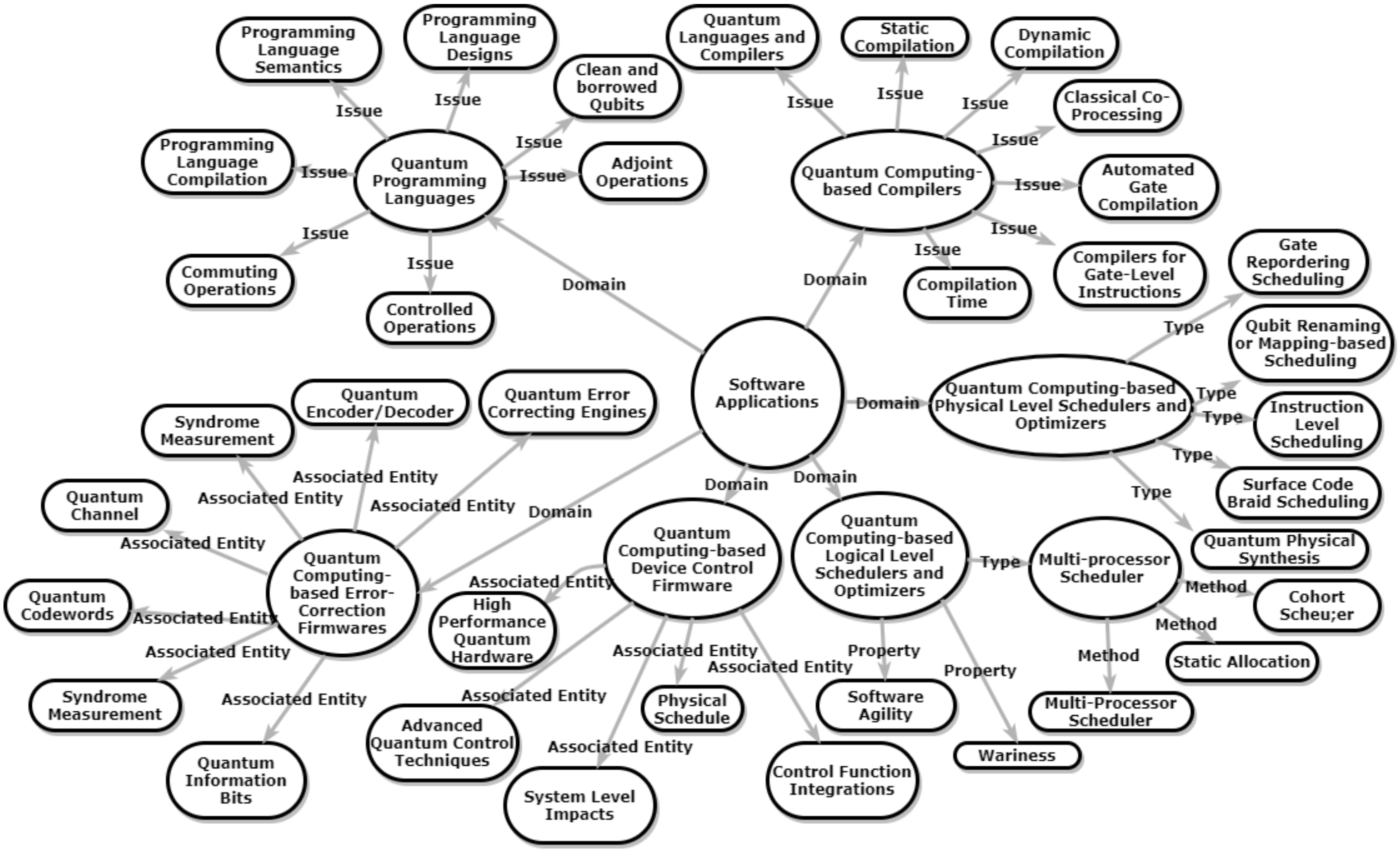

**Figure 7:** Software Applications in Quantum Computing

- ***Quantum Computing-based Compilers:*** In compilation, the important areas explored in recent studies include (i) Quantum Languages and Compilers (QLC), Static Compilation (SC), Dynamic Compilation (DC), Classical Co-Processing (CC), Automated Gate Compilation (AGC), Compilers for Gate-Level Instructions (CGLI), and Compilation Time (CT). According to Chong et al. [270], the major challenges in designing quantum computing-based compilers include (i) designing a hybrid system that supports the compilation of algorithms to gate and machine level instructions,(ii) time, memory, and cost-optimized compilation time, (iii) capability to ensure parallelism and optimal scheduling operations for practical scenarios, and (iv) coordinated compilation between quantum and classical processing. In this coordination, classical processing communicates the precision requirements whereas quantum computation communicates the noise and effort information in hybrid systems.

- ***Quantum Computing-based Logical Level Schedulers and Optimizers*****:** In [271] [272] [273] [274] [275] [276], logical scheduling and optimization studies are analyzed in quantum computing-related scenarios. For example, Oskin et al. [271] discussed the role of dynamic quantum compiler/scheduler in fault-tolerant quantum computing architecture. The processor used in fault-tolerant quantum computing architecture takes logical quantum operations in addition to other control flows and qubit operations. The major challenges in this area include the execution of all quantum algorithms with error correction approaches which make the whole architecture inefficient. Thus, performance optimization, priority-based error measurement, and knowledge of dynamic compilation and algorithm execution should be studied. Likewise, various challenges associated with logical level schedulers and optimizers include (i) integrating the use of multi-processor schedulers and optimizers for error-free quantum physical synthesis, cohort scheduling, and scenarios, where there is a need to apply static compiler/processor allocation, (ii) compilation time, is another important challenge in the quantum world. Thus, applying software agility processes to optimize the job execution with the integration of heavy quantum mechanics (like error-correcting codes with all quantum algorithms) is important to consider, and (iii) lack of trust (wariness) in quantum nodes of quantum network raises concerns over fair and transparent scheduling and optimization jobs. Thus, authentic nodes should be considered to assign the scheduling jobs. Trusted and authentic node identification for scheduling and optimization is important to consider in the future.

- ***Quantum Computing-based Physical Level Schedulers and Optimizers:*** Physical synthesis, scheduling, and optimization processes are important to reduce the latency in quantum circuits, improve the performances, proper circuit allocation, and efficient sharing of the resources between processes. In this process, the important challenges that need to be addressed include [277] [278] [279]: (i) how to apply proper placement and routing heuristics in physical design layout, (ii) to design effective data flow-based gates or circuit placement and routing. Various students in recent times have explored graph-based data flow approaches to accomplish this task. (iii) to apply proper instruction-eel scheduling in instruction issue logic to quantum gates and circuits, (iv) to apply iteration of optimization loops in the scheduling information and incremental updates in scheduling processes, and (v) among other challenges, identification of appropriate heuristic algorithm, error analysis approach, and performance analysis (e.g. time complexity analysis) are required to be studied in future.

- ***Quantum Computing-based Error-Correction Firmwares/Software:*** An efficient quantum error-correction firmware integrates the quantum algorithms and imperfect hardware efficiently. Error-correcting quantum firmware lies at the lowest level of the quantum computing stack and helps in reducing the error caused by imperfect hardware, its complexity, and resource intensity. In quantum error-correcting codes, the important direction to explore include (i) designing efficient quantum error-correcting engines, (ii) developing high-performance quantum hardware, (iii) apply advanced quantum control techniques to operate the hardware, (iv) effectively handle the quantum information bits in storage, processing, and transmission stages, (v) apply appropriate syndrome measurement approach, (vi) integrating all algorithms with quantum codewords, (vii) to apply an effective quantum error-correcting approach that supports quantum channel (with quantum and classical information processing), (viii) to integrate high-quality error encoder and decoder at two ends of data transmission.

- **Quantum Computing-based Device *Control Firmware:*** Software that handles the quantum hardware are expected to provide high performance, ability to apply advanced quantum control techniques, high-quality system-level impacts, simulation-optimization based control for local and global optimum solutions, and appropriate physical schedules.

- ***Quantum Annealing:*** Quantum annealing helps in the identification of the global minimum of a given objective function and set of possible solutions. Quantum annealing is used in various software-based components and applications [280] [281] [282] [283]. Figure 8 shows the usage and associated entities of quantum annealing observed in recent studies. In recent studies [280] [281] [282], quantum annealing is found to be applied in developing a system that automatically reduces the stress level, factoring the pseudo-random functions, analyzing the cybersecurity data, and other applications including finance and healthcare data analysis [284]. Quantum annealing can be used like simulation annealing to find optimum solutions using well-defined single or multi-objective functions. Thus, can be used to handle various problems like the Max-Flow problem in a quantum computer [285]. Further, quantum annealing can be used in error correction codes and software architectures to find the global minimum solution in a different set of problems.

- ***Quantum Software Engineering:*** In [286] [287], recent trends of quantum software engineering are studies. It has been observed that quantum software engineering includes the domain like quantum programming, quantum software tools, quantum software modeling, quantum software implementation, quantum software life cycle, quantum software processes, and quantum software quality attributes. Additionally, this domain studies the syntax and semantics used in programming languages to develop software, quantum, and dynamic logics applied in assertional reasoning to solve challenges in software, characterizing the contract-based disciplines to quantum software.

- ***Quantum Software Testing:*** Like classical software testing techniques, quantum software testing also includes various domains like quantum fuzz testing, quantum functional, white-box, and model-based testing, quantum program debugging, quantum debugging tactics, bug identification, and quantum software verification and validation processes.

- ***Priority Areas:*** In quantum computing, certain priority areas are focused largely to make quantum computers and computing a reality. Few examples of such areas include quantum hybrid systems (supporting classical and quantum computation together), quantum program testing (for verifying and validating the program outcomes), quantum program quality assurance, quantum re-engineering and modernization, quantum software maintenance, quantum software reuse, quantum pattern analysis (using quantum artificial intelligence or quantum machines learning), and quantum circuit analysis.

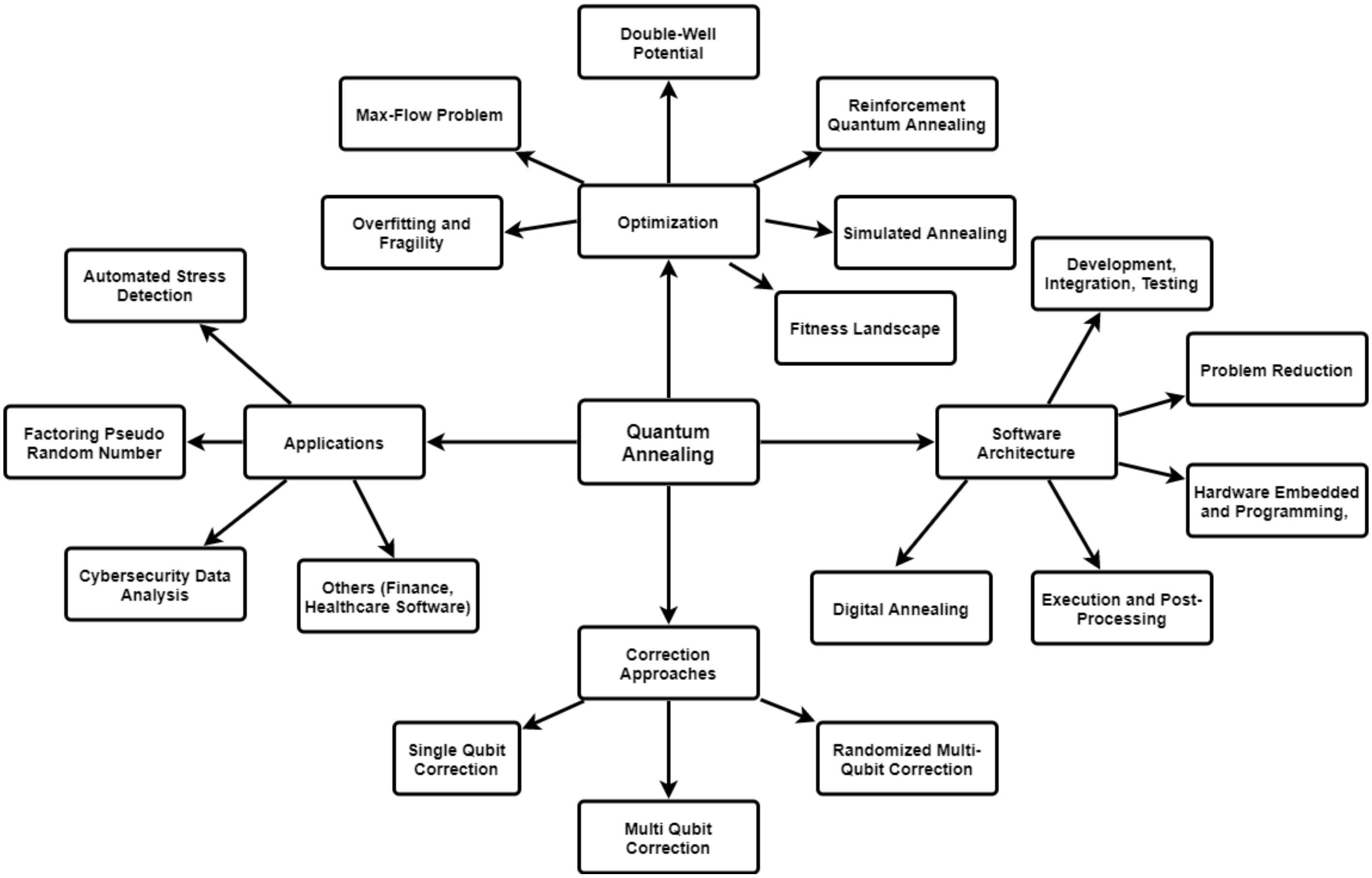

**Figure 8:** Quantum Annealing-based Software Components

- *Other Software Dimensions:* In addition to the above-discussed domains, there are a large set of quantum software aspects that can be explored [288]. For example, quantum provenance in the quantum circuit, data analysis, and quantum compilers. Figure 9 shows the various software-related entities in the quantum software life cycle that need exploration in detail.

## 5. QUANTUM AND POST-QUANTUM CRYPTOGRAPHY

Quantum and Post-quantum cryptosystems are the completely independent world. Quantum cryptography is applying quantum mechanics to perform cryptography tasks. Quantum cryptography encrypts data at the physical network layer by using quantum mechanics' physics. Whereas, post-quantum cryptography or quantum-resistant cryptography uses mathematical techniques. These techniques depend on hard arithmetic problems, which quantum computers cannot answer. Post-quantum cryptography usually refers to algorithms that have the capabilities to secure against attacks. For example, quantum computers running Shor's algorithm can break the security of most standard and difficult-to-solve mathematical cryptography problems [17]-[19]. Likewise, factoring and discrete logarithms, which are widely used in classical cryptosystems, can be solved efficiently on quantum computers. Thus, quantum computing or processing has brought fundamental challenges to the classical cryptosystem. The major applications of quantum cryptography include dense coding, teleportation, prime factorization, faster and secure database searching, secure secret sharing, secure processing, secure one-to-one communication, secure communications across public networks using a quantum smart card and security for cloud and e-commerce computing environments. Quantum computers try every possible solution at same time. Likewise, quantum computers, like brute force attacks, attempt all possible solutions to classical cryptography challenges simultaneously. Thus, regardless of key length, the future of Data Encryption Standard (DES) and Rivest–Shamir–Adleman (RSA) is bleak. Thus, it is important to address whether a quantum machine defend itself against a quantum machine attack or not? In case of unlimited quantum key length, quantum cryptography is considered to be secure. One-time pad is an example of unlimited key length crypto-system. Quantum Key Distribution (QKD) is based on one-time pad. Thus, it is assumed to be secure. There are various ways to ensure secure QKD. For example, Quantum Drones (QD) and Quantum Satellites (QS) are recently explored to share keys and establish multimedia communications [257][258]. Additionally, there are many challenges that are yet to address [259]. For example, (i) quantum nodes are indistinguishable entity in quantum networks. Quantum nodes can be a quantum a repeater, access node or central control node. The major challenge in quantum node is the need to create an efficient buffer mechanism for storing key and meet the dynamic need of quantum network, (ii) to establish an efficient quantum link between nodes with higher key generation rate and lesser cost, (iii) to design and implement a hybrid network consisting of trusted nodes and active optical switches for direct peer-to-peer channel establishment between any two quantum nodes, (iv) to explore the feasibilities of point-to multipoint or single receiver and multiple distribution mechanisms in quantum information processing, (v) to make efficient and trusted quantum multi-path strategies for quantum information processing and distribution, and (vi) to make secure interface between classical system and quantum node for integrating the quantum networks with classical systems.

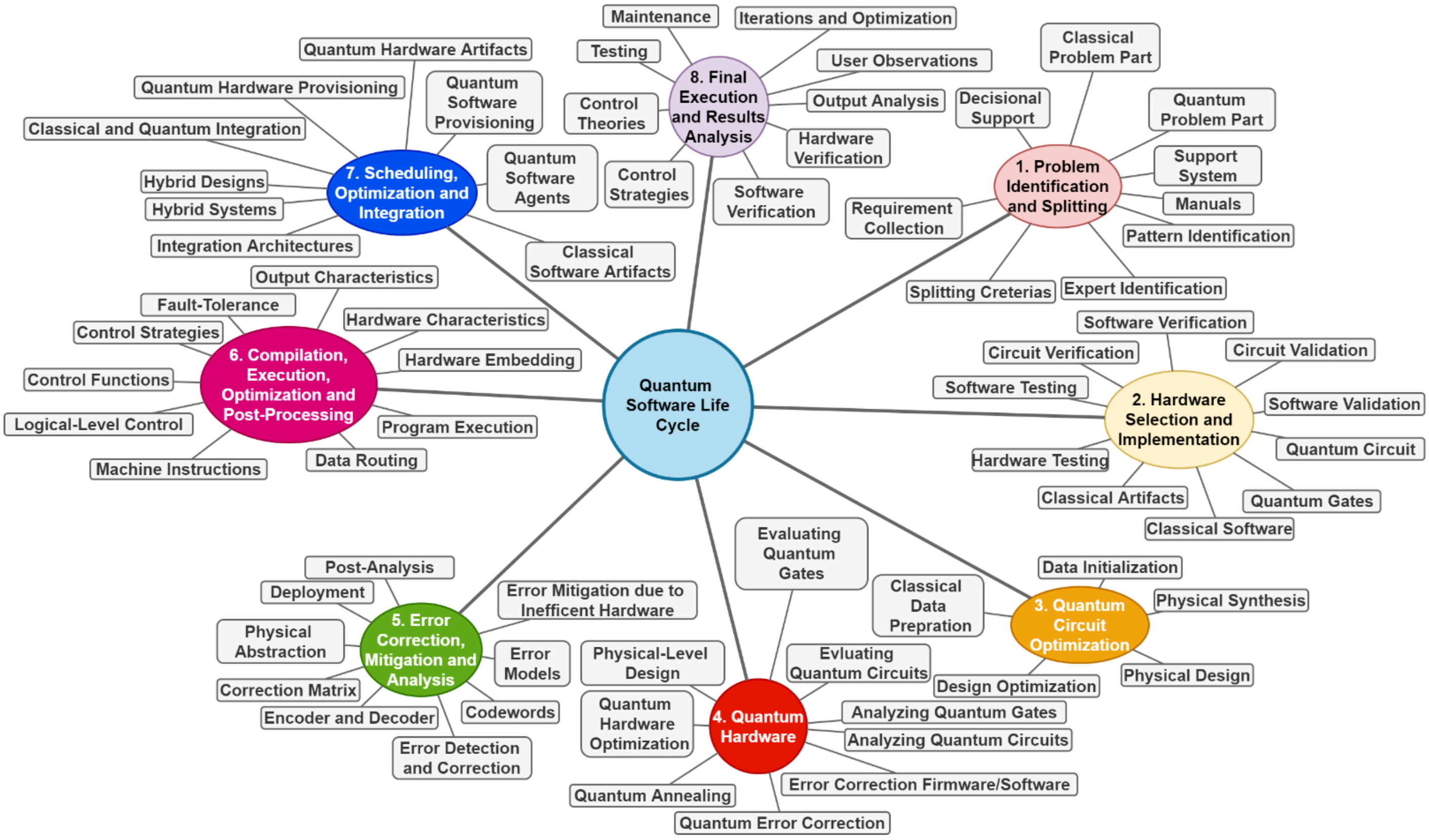

Figure 9: Quantum Software Life Cycle and Associated Terminologies

Post-quantum cryptosystem considers the presence of quantum adversary's challenges due to unique quantum computing features such as no-cloning [19]. Quantum cryptography is defined as quantum mechanical properties for cryptography tasks such as quantum key distribution, encryption/decryption, signature, authentication, and hashing [17] [18] [19] [118]. The major advantages of quantum cryptography include the usage of fundamental laws of physics rather than mathematics-based algorithms which are simple to use but counterintuitive and consume fewer resources. Post-quantum cryptography can be used in various government applications to ensure secure identity proofs. For example, identity-based applications and documents (epassport, national identity cards, and other travel documents) can be made secure with digital signature and encryption processes from quantum-attacks. Post-quantum cryptography can be used in information and communication technologies including networks, networking equipments, servers and network services (e.g. cloud services). Thus, it is an efficient tool for securing futuristic networks. Bringing post-quantum cryptography in automation world can lead to security in various futuristic applications like robotics in healthcare, autonomous vehicles (ground, aerial and underwater), agriculture, and aviation. The major challenges that post-quantum cryptography need to address in future include (i) limit the size of encryption keys or keys used in signature without compromising over security, (ii) the encryption or decryption mechanisms are required to be time efficient for each quantum network entity including quantum communication channel or quantum node, (iii) to reduce the amount of traffic in encryption/decryption or signature processes, (iv) to make an era of quantum computing, quantum algorithm, quantum tool and techniques, quantum technologies, and mathematical standards for speed-up the security scenarios, (v) to provide high bandwidth possibilities for existing network infrastructure and architecture for handling the high traffic scenarios due to post-quantum approaches, and (vi) copying quantum state's encoded data is not feasible, and this reduces the chances of attack and increases the probability of eavesdropping detection, better performance as compared to traditional cryptography etc. Thus, designing efficient buffer-based quantum network device need to be taken in future. Table 5 shows a comparative analysis of quantum cryptography approaches designed and experimented with during recent times. These approaches are classified based on various parameters including designed or experimented for communication protocols, implementation, simulation, quantum-based authentication mechanisms, quantum-based encryption/decryption operations, quantum key distribution, quantum attack detection & analysis, short survey, long survey, approaches using programming for quantum operations, long or short-distance entanglement, attacks (efficiency-mismatch, detector-blinding, detector dead-time, beam-splitter, spatial-mode, eavesdropping), and approaches where data analysis is performed either using machine learning.

Figure 10 shows the classification of quantum cryptography challenges. The challenges are categorized into four major categories, including security attacks and challenges, hardware challenges, performance and cost-related challenges, and quantum-related design challenges. The majority of security attacks and challenges considered various types of security attacks and their feasibility in the quantum world; hardware challenges include experimentation issues whose performance is affected by the hardware used. Performance and cost-related challenges include reducing the cost while improving the performance parameters. Finally, design challenges include developing novel quantum protocols, tools, or techniques while addressing the challenges of existing real-time experimentations.

Table 5: Quantum Cryptography Approaches

| Author | Year | A | B | C | D | E | F | G | H | I | J | K | L | M | N | O | P | Q | R | S |
|---|---|---|---|---|---|---|---|---|---|---|---|---|---|---|---|---|---|---|---|---|
| Deutsch et al. [119] | 1996 | ✔ | ✗ | ✗ | ✗ | ✗ | ✗ | ✔ | ✗ | ✗ | ✗ | ✗ | ✗ | ✗ | ✗ | ✗ | ✗ | ✗ | ✔ | ✗ |
| Naik et al. [120] | 2000 | ✔ | ✗ | ✔ | ✗ | ✗ | ✔ | ✗ | ✗ | ✗ | ✗ | ✗ | ✔ | ✗ | ✗ | ✗ | ✗ | ✗ | ✔ | ✗ |
| Elboukhari et al. [121] | 2010 | ✔ | ✗ | ✗ | ✗ | ✗ | ✔ | ✗ | ✔ | ✗ | ✗ | ✗ | ✗ | ✗ | ✗ | ✗ | ✗ | ✗ | ✗ | ✗ |
| Bugge et al. [122] | 2014 | ✔ | ✔ | ✗ | ✗ | ✗ | ✔ | ✔ | ✗ | ✗ | ✗ | ✗ | ✗ | ✗ | ✗ | ✗ | ✔ | ✗ | ✔ | ✗ |
| Jain et al. [123] | 2014 | ✔ | ✔ | ✗ | ✗ | ✗ | ✔ | ✔ | ✗ | ✗ | ✗ | ✗ | ✗ | ✗ | ✗ | ✗ | ✗ | ✗ | ✔ | ✗ |
| Bruss et al. [124] | 2017 | ✔ | ✗ | ✗ | ✔ | ✔ | ✔ | ✔ | ✗ | ✔ | ✗ | ✗ | ✗ | ✗ | ✗ | ✗ | ✗ | ✔ | ✔ | ✗ |
| Li et al. [125] | 2018 | ✗ | ✗ | ✗ | ✔ | ✔ | ✔ | ✔ | ✔ | ✗ | ✗ | ✗ | ✗ | ✗ | ✗ | ✗ | ✗ | ✔ | ✔ | ✗ |
| Bennett and Brassard [126] | 2020 | ✔ | ✗ | ✗ | ✗ | ✗ | ✔ | ✗ | ✗ | ✗ | ✗ | ✗ | ✗ | ✗ | ✗ | ✗ | ✗ | ✗ | ✗ | ✗ |
| Bhusal et al. [127] | 2020 | ✔ | ✗ | ✔ | ✗ | ✔ | ✗ | ✗ | ✔ | ✗ | ✗ | ✗ | ✗ | ✗ | ✗ | ✗ | ✗ | ✔ | ✔ | ✔ |
| Brassard et al. [128] | 2000 | ✔ | ✗ | ✗ | ✗ | ✗ | ✔ | ✔ | ✔ | ✗ | ✗ | ✗ | ✗ | ✗ | ✗ | ✗ | ✗ | ✗ | ✔ | ✗ |
| Durak and Jam [129] | 2020 | ✔ | ✔ | ✗ | ✔ | ✗ | ✔ | ✔ | ✗ | ✗ | ✗ | ✗ | ✗ | ✔ | ✗ | ✗ | ✗ | ✗ | ✗ | ✔ |
| Gras et al. [130] | 2020 | ✔ | ✔ | ✗ | ✗ | ✗ | ✔ | ✗ | ✗ | ✗ | ✗ | ✗ | ✗ | ✗ | ✔ | ✔ | ✗ | ✗ | ✗ | ✗ |
| Guo et al. [131] | 2020 | ✔ | ✗ | ✔ | ✗ | ✗ | ✔ | ✔ | ✗ | ✗ | ✗ | ✗ | ✗ | ✗ | ✗ | ✗ | ✔ | ✔ | ✔ | ✗ |
| Huang et al. [132] | 2020 | ✔ | ✔ | ✗ | ✗ | ✗ | ✔ | ✔ | ✗ | ✗ | ✗ | ✗ | ✗ | ✗ | ✗ | ✗ | ✗ | ✗ | ✔ | ✗ |
| Melhem et al. [133] | | ✗ | ✗ | ✗ | ✗ | ✔ | ✔ | ✔ | ✗ | ✗ | ✗ | ✗ | ✗ | ✗ | ✗ | ✗ | ✗ | ✗ | ✗ | ✗ |
| Qi et. al. [134] | 2020 | ✔ | ✔ | ✗ | ✔ | ✔ | ✔ | ✗ | ✗ | ✗ | ✗ | ✔ | ✔ | ✔ | ✗ | ✔ | ✗ | ✔ | ✔ | ✗ |
| Shang et al. [135] | 2020 | ✔ | ✗ | ✗ | ✔ | ✗ | ✔ | ✔ | ✗ | ✗ | ✗ | ✗ | ✗ | ✗ | ✗ | ✗ | ✗ | ✗ | ✔ | ✗ |
| Trushechkin et al. [136] | 2020 | ✔ | ✗ | ✗ | ✗ | ✗ | ✔ | ✗ | ✗ | ✗ | ✗ | ✗ | ✗ | ✔ | ✗ | ✗ | ✗ | ✗ | ✗ | ✗ |
| Vybornyi et al. [137] | 2020 | ✔ | ✗ | ✗ | ✗ | ✔ | ✔ | ✔ | ✗ | ✗ | ✗ | ✗ | ✗ | ✗ | ✗ | ✗ | ✗ | ✗ | ✔ | ✗ |
| Yin et al. [138] | 2020 | ✗ | ✔ | ✗ | ✔ | ✔ | ✔ | ✔ | ✗ | ✗ | ✗ | ✔ | ✗ | ✔ | ✔ | ✔ | ✔ | ✔ | ✔ | ✗ |
| Zhang et al. [139] | 2020 | ✔ | ✗ | ✔ | ✗ | ✗ | ✔ | ✔ | ✗ | ✗ | ✗ | ✗ | ✗ | ✔ | ✗ | ✗ | ✗ | ✗ | ✗ | ✔ |
| Zhou et al. [140] | 2020 | ✔ | ✔ | ✗ | ✗ | ✗ | ✔ | ✗ | ✗ | ✗ | ✗ | ✗ | ✔ | ✗ | ✗ | ✗ | ✗ | ✗ | ✗ | ✔ |

A: Communication protocols, B: Implementation, C: Simulation, D: Quantum Authentication, E: Quantum Encryption, F: Quantum key distribution, G: Quantum Attack Detection & Analysis, H: Short survey associated with implementation , I: Long Survey for in-depth analysis , J: Quantum Programming, K: Long-distance entanglement, L: Short-distance entanglement, M: Efficiency-mismatch attack, N: Detector-blinding attack, O:Detector dead-time attack, P: Beam-splitter attack, Q: Spatial-mode attack, R: Eavesdropping attack, S: Data analysis/Machine Learning

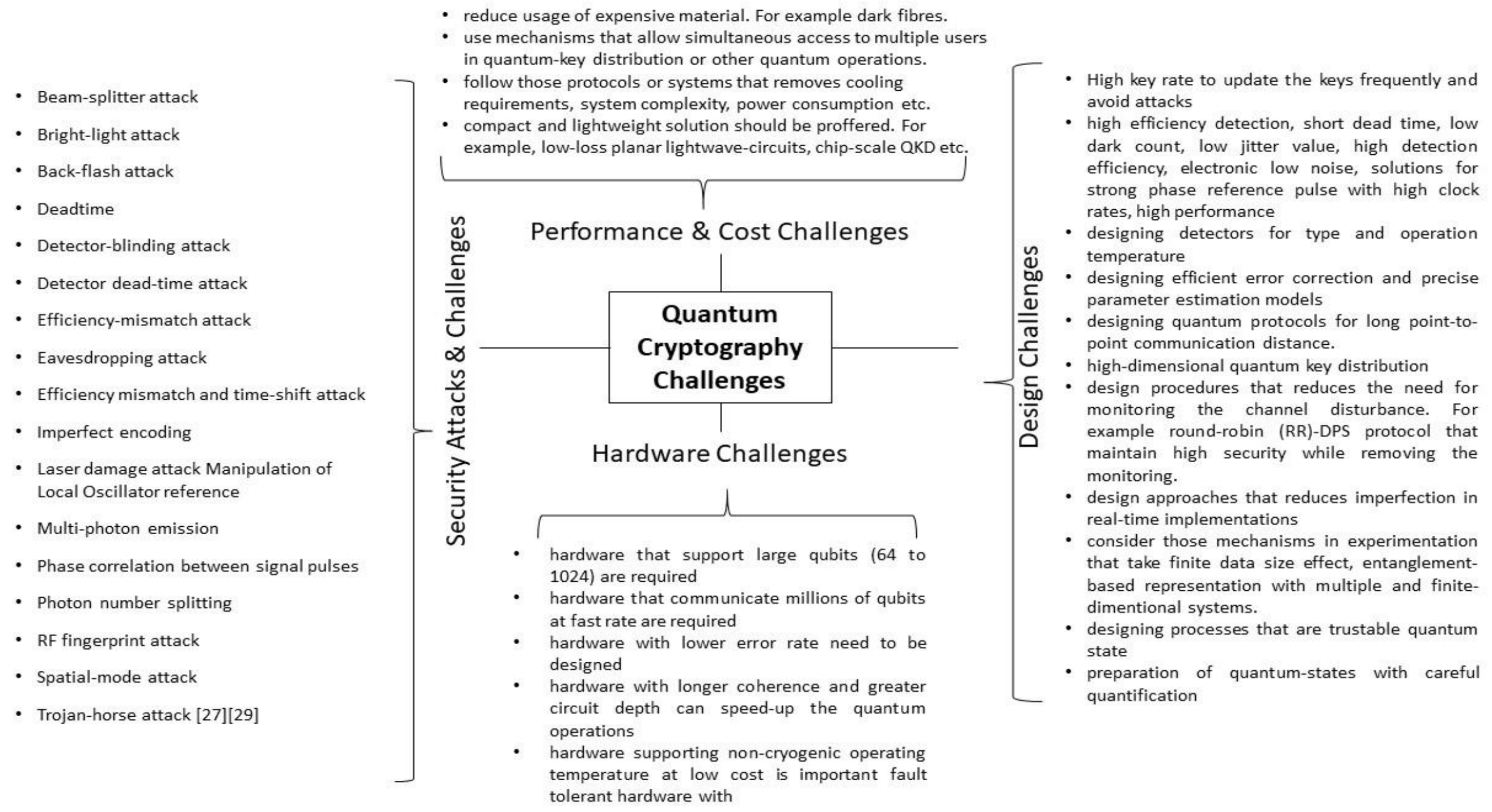

Figure 10: Quantum Cryptography Challenges

## 5.1 QUANTUM KEY DISTRIBUTION (QKD)

Quantum Key Distribution (QKD) is an effective way of protecting information security using quantum computers [49] [141]. As compared to traditional cryptography-based key distribution mechanisms, which are vulnerable to computational power-based scenarios, a quantum cryptography mechanism (like QKD) is secure against various attacks. In quantum cryptography, the no-cloning theorem [142] in quantum mechanics states that it is impossible to make a perfect copy of the quantum system or its states. Thus, any eavesdropping attempt adds noise to the quantum transmission that is easily detectable by two parties (source and destination) [143]. QKD protocols can be classified based on the use of properties during transmission including applied modulation, encoding/decoding, and quantum channel implementation. Likewise, there are various types of quantum key distribution approaches [144] [145]. Table 6 shows the comparative analysis of these approaches, which are discussed below.

**Table 6:** Comparative analysis of QKD

| Type of CV-QKD | Pros | Cons |
|---|---|---|
| Gaussian-modulated CV-QKD | • Security analysis is much more advanced compared to discrete-modulated CV-QKD. | • Distance limitation for secure QKD is a major concern.<br>• The use of high-performance error-correcting code can improve security but reduces the distance coverage [148]. |
| Discrete-modulated CV-QKD | • More suitable for long-distance secure key transmission.<br>• Simple experimentation setup.<br>• Great potential for large-scale deployment in secure quantum networks.<br>• The integration of post-selection strategies with reverse reconciliation can significantly improve the key rates. | • Security analysis in this system is more challenging compared to Gaussian-modulated CV QKD because analysis relies on the linearity of the channels which is not an easy condition for verification. |
| Coherent On-Way (COW) Quantum Key Distribution | • Simple in experimentation<br>• Reduce interference visibility<br>• Avoid photon number splitting attack to a large extent<br>• Falls in distributed-phase-reference QKD category | • Empty pulses contain a light that can introduce noise. This can increase error rates.<br>• Performance decreases with an increase in disturbances. Small disturbances do not affect performance. |
| Differential Phase-Shift (DPS) Quantum Key Distribution | • Falls in distributed-phase-reference QKD category.<br>• Integration with randomness or improved transmitter can reduce the disturbances and improve the performance [151] [152]. | • Chances of side-channel attacks are higher. Thus, techniques (e.g. attenuation) are required to be integrated for removing it.<br>• Performance decreases with an increase in disturbances. Small disturbances do not affect performance. |
| Six-State Quantum Key Distribution | • Using this category of protocols, a high error-rate can be detected easily in the presence of any eavesdropping attack.<br>• The speed of communication lies in the high-speed key distribution category.<br>• The probability of interference, collective attack, and obtaining the secret is low. | • Chances of obtaining the secret cannot be completely avoided.<br>• Need to analyze the hidden variable models for protecting the protocols against attacks.<br>• Multiple eavesdropping challenges to topple authenticated communication need to be addressed. |
| Decoy-State Quantum Key Distribution | • A high secure key rate can be generated using the decoy-state protocol.<br>• The problem of lower secure key rate can be efficiently handled with inequality based statistical models.<br>• Found to be an effective method in avoiding the photon-number splitting attack. | • The secure key rate can be lower down to a significant level if parties' parameters are varied with different decoy states.<br>• Usage of different decoy states is not yet experimented in realistic scenarios to confirm its adaptability with security.<br>• Computational power challenge also reduces the secure key rate and can cause statistical fluctuations. |

### 5.1.1 DISCRETE AND CONTINUOUS VARIABLE QUANTUM KEY DISTRIBUTION

QKD can be designed in both discrete and continuous variables. Some of these approaches are briefly discussed as follows. Ghalaii et al. [142] discussed the Gaussian and non-Gaussian modulated Continuous-Variable QKD (CV-QKD) methods. Further, the non-Gaussian CV-QKD protocol is extended with a discrete modulation approach for increasing the secret key rates. Besides, the proposed mechanism is found to support the discrete-modulation CV-QKD over CV quantum repeaters and to long-range system operation in-live. Valivarthi et al. [146] proposed a plug-and-play CV-QKD with Gaussian modulation quadratures. In experimentation, two independent fiber stands have been used for two narrow line-width lasers for quantum signal transmission. This experimentation increases the secret key rate up to 0.88 Mb/s with different experimental setups and inputs. This experimentation is considered to be an effective mechanism in terms of low-cost deployment for metropolitan optical networks. The investigation is useful in terms of its design, use of Raleigh back-scattering mechanism to minimize noise, and integration of GG02 symmetric protocol with heterodyne detection [143]. The complete setup makes the proposed QKD faster and secure. Leverrier and Grangier presents a CV-QKD protocol combining discrete modulation and reverses reconciliation [147]. The protocol is tested experimentally, and it is observed that the proposed scheme can distribute the secret key over a long distance while ensuring security. Li et al. [148] proposed a discrete modulated CV-QKD scheme that improves the system performance and secure distance with machine-learning-based detectors. The proposed scheme is capable of processing the secret keys to improve the overall system performance. Lin et al. [149] applied numerical methods to analyze the security aspects in discrete modulated CV-QKD. The two proposed variants of discrete-modulated CV-QKD are capable of generating much high key rates for longer distances as compared to binary or ternary modulation schemes. Thus, aim of this approach is to generate high key rates for longer distances as well. Ruan et al. [150] analyzed optical absorption and scattering properties of discrete-modulated CV-QKD. It is observed that the performance of four and eight-state protocol in asymptotic and finite-size cases is dependent on seawater composition, i.e. if the composition is complex, then the performance of protocol decreases as well. The variation in optical modulation and minimizing the extra noise can improve the protocol's performance. In another observation, it was found that the number of states improves performance. In this case, the performance of the eight-state protocol is better compared to the four-state protocol. In recommendations, CV-QKD is found to be significant over the seawater channel and provides a good medium to construct a secure communication network.

### 5.1.2 COHERENT ON-WAY (COW) QUANTUM KEY DISTRIBUTION

In COW QKD, logical bits are encoded and emitted with a sequence of weak coherent pulses. These pulses may be tailored from a laser having an intensity modulator. Various mechanisms are used to make COW QKD practical. Some of the recently discussed approaches are explained as follows. Stucki et al. [153] presented a COW-QKD protocol with weak coherent pulses. The simplicity of this experimentation increases the bit rates and reduces interference visibility as well. This protocol achieves a high efficiency for secret bits per qubit generation while lowering the photon number splitting attacks. Mafu et al. [154][155] realized the importance of the differential-phase-reference category of QKD protocols. In this category, there are mainly two types of QKD protocols, including COW and differential phase shift. Mafu et al. [155] formalized the COW-QKD protocol with non-computing Positive Operator-Value Measures (POVM). This formalization increases the chances to have unconditional security proof against general attacks. The POVM elements, effective for generating security proofs against attack, include measurement probabilities and positive operators, composite measures, all types of measurements distinguishing between two quantum states, and an informationally complete secure state. Mauf et al. [159] reanalyzed the necessary condition requirements with non-commuting POVM elements-based COW protocol. The major challenge considered in stating unconditional security proof is the class of protocol that uses coherent signals. These coherent signals are not symmetric as compared to qubits used in proof realization. Thus, there is a need to formalize the COW QKD protocol without disclosing the detailed working explanations and parties' confidentiality. It is observed that POVA elements can make this possible with high-security standards. Wonfor et al. [157] conducted a trial of the COW-QKD protocol with a commercial-grade encrypted system. In experimentation, a link is launched for QKD with 500 Gbps encrypted data transmitted over a distance of 121 Km. As result, it is observed that QKD in O-band COW protocol with free detectors and C-band DWDM channels gives a stable performance for many weeks. Further, 25 DWDM channels with co-propagation can make the QKD process feasible while ensuring security proofs.

### 5.1.3 DIFFERENTIAL PHASE-SHIFT (DPS) QUANTUM KEY DISTRIBUTION

In DPS QKD, a highly attenuated coherent pulse with phase shift is sent from the sender side and is received with a one-bit delay at the receiver side. It is a long since this approach was developed. However, several variations of this approach are studied in recent times. Some of these approaches are discussed as follows. Alhussein and Inoue [158] realized the importance of side-channel attacks in the DPS-QKD system. DPS-QKD protocol is found to be another simple and efficient protocol because it works in cases when precise synchronization of signals between distant parties is not possible. The proposed scheme has avoided the control of blinding and controlling side-channel attacks. To detect a side-channel attack at Bob's side, a variable attenuator is added at random and occasional attenuation inserted. Further, the performance is analyzed, confirming the adaptability of the proposed approach. Collins et al. [159] experimented with the quantum digital signatures transmission over a long distance (90 Km) using the DPS-QKD protocol. The authors claimed that the transmission was aimed to be conducted for long-distance compared to previous works. The distribution of quantum digital signatures ensures message integrity as well as non-repudiation. Further, the performance of the proposed scheme is comparable to the BB84 protocol used for QKD with 1550 nm wavelength and similar experiment settings, including clock rate and transmission distance considered for the operation. Hatakeyama et al. [151] experimented with a round-robin DPS-QKD protocol to reduce the bit error rates. The experiment is conducted to take advantage of simple DPS-QKD functioning to increase tolerance without compromising on security issues. This work has extended with basic DPS-QKD protocol with randomness. The randomness and few additional delays increase the performance of the proposed protocol as compared to the basic DPS-QKD protocol. The simulated experimentation and key generation rates are analyzed with different randomness patterns. It is observed that the performance of the proposed protocol can be significantly increased with a few parameter changes. Schrenk et al. [152] developed a low-complexity transmitter for DPS-QKD. This transmitter uses an integrated laser device with two electro-optic elements. This experimentation observed the quantum state preparation and chances of side-channel attacks with the proposed transmitter mechanism. A distributed environment with a centralized quantum receiver shows the performance of form-factor and successful deployment at a short-term distance. Overall, the performance of the whole system is found to be effective for QKD compared to generalize DPS-QKD. Sibson et al. [160] identified a low error rate; high speed clocked QKD operation of indium phosphide transmitter chip useful in the telecommunications industry. This configuration has experimented with three protocols, including BB84, coherent one way, and different phase shifts. Results show that the proposed approach gives better performance without impacting the security standards, and they are useful for any sort of communications in telecommunication networks.

### 5.1.4 SIX-STATE QUANTUM KEY DISTRIBUTION

In six-state quantum cryptography protocols, BB84 protocol is extended to use six-state polarization (|0>, |1>, |+i>, |-i>,|+>, |->) on three orthogonal bases. Further, the six-state protocol can tolerate a noisier channel and detect higher rate errors during any eavesdropping attack. The six-state protocol can be implemented either using a quantum computer or optical technologies. For example, Lo [161] derived the proofs for unconditional security solutions in six-state quantum key distribution protocols. In this implementation, it has been observed that unconditional security could lie at a high bit error rate of 12.7% as compared to 11% in the BB84 protocol. The proposed technique has used DiVincenzo, Shor, and Smolin’s quantum codes for bit-flip and phase error pattern analysis. It has been observed that bit-flip error syndromes entropy can be used for a phase error pattern that increases the security of the proposed protocol at a high error rate as well. Similarly, Azuma et al. [162] realized the security of the six-state quantum key distribution protocol against various attacks, including intercept/resend, collective, and eavesdropping. Here, the probability of an attacker’s interference in legitimate user communication is noticed, and the chance of obtaining the secret is measured. In collective attack observations, the security level is found to be high that can protect imposing looser constraints upon the attacker’s strategies. This work has considered the comparative analysis of proposed security-level detection with the E91 protocol. Results show that the six-state protocol is comparatively secure against attacks if hidden variable theories are examined with a small

disturbance of 1/3. Chau et al. [163] identified that four-dimensional qubits in quantum key distribution are possible, and it can have security equivalent to the six-state scheme with arbitrarily long raw key size. Here, the tolerance level is observed to be 21.6% using one-way classical communication with passive basis selection in decoy. Thus, an increase in security level with a high key rate meets the requirements of the current quantum key distribution.

### 5.1.5 DECOY-STATE QUANTUM KEY DISTRIBUTION

The decoy-state quantum key distribution protocol is preferred over others because it provides better conditional or unconditional constraints over the gain and the error rate of single-photon states. In recent times, various amendments are made to improve the decoy-state quantum key distribution protocol. For example, Liu et al. [164] realized the importance of decoy-state QKD protocol and its capability to protect against photon-number splitting attacks. In this work, two-basis detector efficiency asymmetry was found to be existing in real experimentation. To improve the rate of QKD with asymmetric basis-detector efficiency asymmetry, this work has investigated a 4-intensity decoy-state optimization protocol to protect against attacks. In observation, it is found that X and Z basis efficiencies are not the same, and the practicality of decoy-state has high chances. Grasselli et al. [165] focused on Twin-Field (TF)-QKD protocol because of a secure secret-key mechanism. It has been observed in an analysis that the security of this protocol is associated with photon-number states using the decoy-state method. This work has derived analytical bounds on the parameters used by parties and concluded that either two, three, or four decoy intensity settings could be used for investigating the protocol's performance. In further observations, the protocol is found to be robust against optical pulses' fluctuations. Chau et al. [166] made various observations in the decoy-state protocol. In the first observation, it is found that a secure key rate can be seriously lowered down with the deviation of single-photon. In their second observation, the error rate can also lower the secure key rate by bounding the yields and usage of the type of decoy. To improve the secure key rate in such conditions, McDiarmid inequality is found to be effective because it helps in computing the lower bound in the centering sequence method. As result, it has been observed that the secure key rate can be doubled with the proposed approach for a realistic 100 km long quantum channel. This work has introduced a powerful inequality technique for handling problems beyond statistical data with the central limit theorem. Liu et al. [167] applied the chernof bound to passive decoy-state and improved the final key rate. In experimentation, it is claimed that the proposed approach can securely transmit the data over 205 km, which is close to an asymptotic limit of 212 km. This is found to be the highest key rate over a long distance compared to existing approaches. In conclusion, the majority of decoy-state protocols are used either to improve the secure key rate or its transmission over a long distance.

## 5.2 POST-QUANTUM CRYPTOGRAPHY

The post-quantum cryptosystem is defined as the set of cryptography primitives and protocols that are secure against quantum computer attacks [47] [48] [168] [169] [170]. It is observed that the existing cryptography primitives and protocols rely on mathematical problems such as integer factorization, discrete logarithm, and elliptic-curve discrete logarithm [51]. With the possibilities of quantum computers, it is theoretically proved that all of these mathematical problems could be solved in a short duration [156]. Thus, post-quantum cryptography is widely discussed. The protocols in post-quantum cryptography are mainly classified into five categories: code-based, lattice-based, supersingular elliptic curve isogeny, multivariate, and hybrid, as shown in Figure 11.

### 5.2.1 LATTICE-BASED CRYPTOSYSTEM

In mathematics, Lattice is an arrangement of regularly spaced points in a subgroup $R^n$ that is isomorphic to another group $Z^n$ such that $R^n$ is isomorphic to $Z^n$ i.e. all combinations of vectors in space lies in $R^n$. Ajtai [171] initiated the use of cryptography in the lattice-based system and derived the computationally hard problems on lattices. The computational hard problem provides security and is found to be useful in other cryptography primitives such as homomorphic cryptography, attribute-based cryptography, and code-based cryptosystem. In [173]-[176], a lattice class of Small Integer Solution (SIS) and its Inhomogeneous variants are discussed. For example, Ring-SIS [173] is a variant of the lattice-based cryptosystem. This variant is having an issue of difficult to solve for any randomly selected instance. This property makes this variant a class of average-case problems and it is not sufficient to have worst-case complexity applications supporting SIS or Ring-SIS variant. Like other lattice-based cryptosystems, SIS and Ring-SIS variant also falls in the NP-hard problem. As compared to a lattice-based system with worst-case hardness, SIS and Ring-SIS are likely to secure against quantum computers. In recent studies, lattice-based schemes are more focused on shifting from worst-case to average-case security perspectives. Thus, breaking the randomly chosen instance in lattice-security schemes and finding a solution for worst-case instances of the lattice-based system are important concerns.

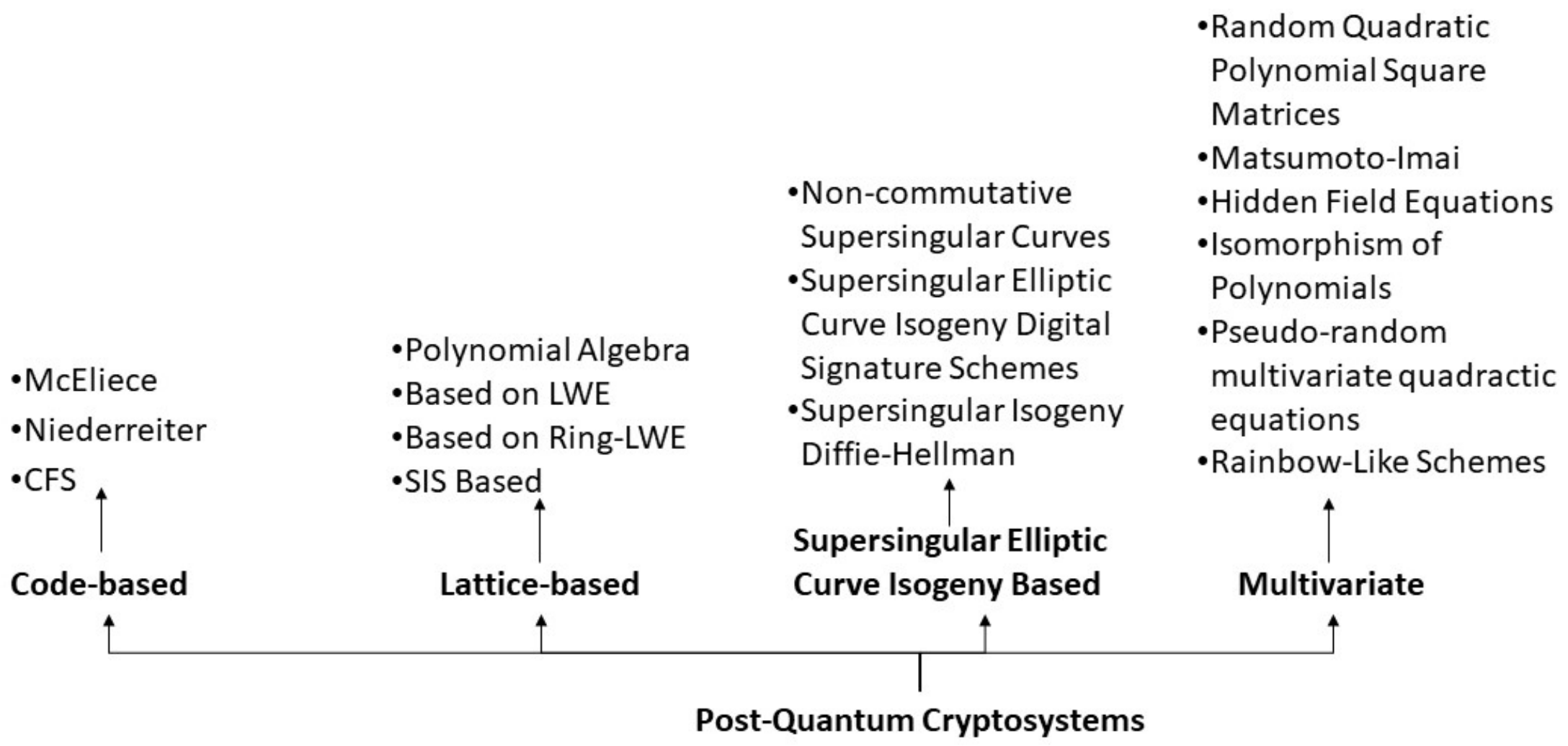

**Figure 11:** Post-quantum Cryptography Protocols

Langlois and Stehlé [289] compared the average-case reduction problems with module lattices in a lattice-based system. The security of both systems is found to be comparable. Cryptosystems based on worst-case to average-case reduction analysis were found to be more secure because of converse reductions. Schemes based on ideal lattices, having structured objects, provide higher security. Assessing those schemes which use ideal lattices to ensure high security need to be assessed to standardize the lattice problems for more general and specific classes of lattices. Exploiting the ideal lattices or module lattices with a certain degree of module rank would impact the lattice-based cryptosystem. The hardness of various schemes (like Ring-SIS, R-LWE) would be impacted with successful ideal lattice exploitation. Plantard and Schneider [290] compared the ideal and general lattices and started the experimentation to create challenges for ideal lattices. It is assumed that the security challenges of lattice-based cryptosystems lie in ideal lattices. Thus, security assessments of ideal lattices are studies in recent work for both SIS and LWE-based lattice cryptosystems. Lyubashevsky et al. [291] explored the ideal lattices in ring signature and confidential transactions. With the use of ideal lattices, the issue of small output, storage and processing can be achieved. Reducing the transaction size and other requirements make lattice-based cryptosystem a viable candidate for resource-constraint networks like IoT networks. The transaction size is an important parameter to consider for employing lattice-based schemes in networks. So far, it is assumed that transaction size depends upon security parameters but efforts are done to reverse this assumption and make it possible for more applications. Various lattice-based cryptosystem approaches are summarized in Table 7 with comparative analysis of their primitives in Table 8.

Table 7: Lattice-based Cryptosystem Approaches

| **Lattice-Problem** | **Variants** | **Pros** | **Cons** |
|---|---|---|---|
| SIS and its Inhomogeneous Variants | Ring-SIS [173], Bi-GISIS [174], Lattice-based Direct Anonymous Attestation (LDAA) [175], Certificateless Signature (CLS) scheme on NTRU lattice [176]. | Smaller storage and faster operations are preferred.<br>Schemes, like LDAA, are secure against weak/strong deniability attacks. | Weak/strong deniability is the least addressed.<br>The SIS problem becomes solvable in polynomial time with various parameter variations. |
| Learning With Errors (LWE) | Ring-LWE [173] [174], MPSign[177], Decision-LWE [175], Decision-Ring-LWE [175], LDAA [175], Module-LWE [178], Module-Learning With Rounding (M-LWR) [179], NewHope [180], Kyber [180], R. EMBLEM [180], KCL [180], OKCN/AKCN-RLWE [180], AKCN-MLWE [180], ILWE [181], MPLWE [181]. | Smaller storage and faster operations are preferred<br>Polynomial-based LWE allows for secrets that are much smaller compared to modulus operations. In results, schemes are faster. | There are equally likely chances of chi-square attack, cyclotomic vulnerabilities, inherent structure exploitability, and sensitive dependence to field parameters in the majority of existing schemes.<br>Weak/strong deniability is the least addressed. |

Table 8: Comparative Analysis of Lattice-based Cryptography Primitives

| Author | Major Observations | Year | A | B | C | D | E | F | G | H |
|---|---|---|---|---|---|---|---|---|---|---|
| Banerjee et al. [178] | A Low-power crypto-processor is designed, configured, and tested to accelerate polynomial arithmetic operations.<br>Lightweight cryptography primitives and protocols are combined with sampling techniques. This accelerates the polynomial sampling in discrete distribution parameters useful in lattice-based schemes. | 2019 | ✔ | ✔ | ✔ | ✔ | ✔ | ✔ | ✔ | ✔ |
| Nejatollahi et al. [181] | Surveyed lattice-based cryptographic schemes, security challenges in software and hardware implementations, and technology adoption | 2019 | ✔ | ✔ | ✔ | ✔ | ✘ | ✔ | ✔ | ✔ |
| Akleylek et al. [174] | An authentication key exchange-based scheme is designed using the Bi-GISIS problem.<br>Comparative analysis with SIS and LWE problems is performed.<br>Testing of the proposed approach with a security model is conducted. | 2020 | ✔ | ✔ | ✔ | ✔ | ✘ | ✘ | ✔ | ✔ |
| Bai et al. [177] | A polynomial LWE-based digital signature scheme is proposed and found to be secure with a quantum-access random oracle model.<br>This work has observed an efficient key-recovery attack against homogeneous polynomial SIS problems with small secrets. | 2020 | ✔ | ✘ | ✔ | ✔ | ✘ | ✘ | ✔ | ✔ |
| El Kassem [175] | In this work, smart zero-knowledge proofs are designed and explained for lattice problems. | 2020 | ✔ | ✔ | ✔ | ✘ | ✘ | ✘ | ✔ | ✔ |
| Mera et al. [179] | Designed and experimented with a polynomial multiplier using the Toom-Cook algorithm for cryptoprocessors in the lattice system.<br>Usage of the proposed hardware-based system is tested for cryptography primitives especially public key protocol. | 2020 | ✔ | ✘ | ✘ | ✔ | ✔ | ✘ | ✔ | ✔ |
| Nejatollahi et al. [180] | This work has explored the design space of a flexible and energy-efficient post-quantum cache-based hardware accelerator for five different submissions. | 2020 | ✔ | ✔ | ✔ | ✔ | ✔ | ✔ | ✔ | ✔ |
| Xu et al. [176] | Proposed quantum attack resilience certificateless signature scheme with the difficulty of small integer solution on the NTRU lattice. | 2020 | ✔ | ✘ | ✔ | ✔ | ✘ | ✘ | ✔ | ✔ |

A: Encryption/Decryption, B: Authentication, C: Digital Signature, D: Key Distribution, E: Cryptoprocessor design, F: Identification Scheme, G: Lattice-based approach for application, H: Protocol Design/Development/Implementation/Simulation,

### 5.2.2 CODE-BASED CRYPTOSYSTEM

Robert McEliece initiated the code-based cryptography based on NP-hardness of the Syndrome Decoding Problem (SDP) [172]. A code-based cryptosystem relies on secretly decoding the linear code having a predefined structure. McEliece scheme is based on binary Goppa codes (as linear code) with the Nicholas Patterson algorithm in the decoding process. McEliece cryptosystem is fast in its encryption and decryption operations. The major drawback of the McEliece cryptosystem is the use of large key sizes that make this scheme infeasible for resource-constrained devices. In literature [182]-[186], various variants of the McEliece scheme are proposed using different error-correcting codes such as Rank ECC, Gabidulin codes, Twisted Gabidulin codes, Twisted Reed-Solomon Codes, Low-Density Parity-Check (LDPC) codes, quasi-cycle codes, and Quasi-Cyclic Low-Rank Parity -Check (QC-LRPC). Among other code-based cryptosystems [186], Niederreiter and CFS (Courtois, Finiasz, Sendrier) cryptosystems are also very popular. The CFS system is found to be useful for Internet of Things (IoT) signature schemes with Fiat-Shamir transformation [187]. Both Niederreiter and CFS schemes generate small signatures that result in fast computations. Table 9 shows the analysis of a few recent contributions in code-based cryptosystems. Various code-based cryptosystem approaches can be classified based on two sets of problems: SDP and the hardness of distinguishing a code from pseudorandom code. These problems and their interconnection with code-based cryptography are explained as follows.

- *Syndrome Decoding Problem*: This is the first set of problems for several code-based algorithms. Over the development of a code-based cryptosystem, the complexity of the syndrome problem is increasing. However, SDP-based code-based cryptosystems share design and complexities. Their complexities rely on Grover or quantum walks. Cayrel et al. [292] discussed the computational efficiencies of linear programming to perform real-time message recovery attacks. This is a message-recovery laser fault injection attack over a code-based cryptosystem. This attack experiments over the classic McEliece Cryptosystem in a worst-case scenario. There are many adversaries or attack feasibilities studied during the recent time in the worst-case scenario. However, an average-case scenario in any post-quantum cryptosystem is considered to be much secure. A reference to this attack to the Niederreiter cryptosystem is also discussed. A large set of code-based cryptosystems and studies are based on either McEleiece or Niederreiter cryptosystems [182]-[186]. The chances of attack increase with more faults in syndrome decoding or fault injection reduction in computational complexity scenarios (like in IoT). With an increase in vulnerabilities, the chances of other attacks like secret error-vector disclosure with efficient linear programming or strong parameter disclosure in cryptosystems. It has been observed that the chances of any form of these attacks increase with variations in the fraction of faulty syndrome entries. Ezerman et al. [185] discussed the hardness of McEliece and the syndrome decoding problem for group signature schemes using code-based cryptography. This scheme applies anonymous and randomness to ensure group signature. Authors have focused on implementing this approach and suggested improving the performance in implementation, applying standard model or quantum random oracle model to get better experiences.

- *The hardness of Distinguishing a Code from Pseudorandom Code*: Rank metrics play important role in various applications associated with code-based cryptography. Few applications include space-time coding, network coding, and asymmetric key-based cryptosystems [293] [294] [295]. Hardness is a theoretical property of a code. In code-based cryptography, the hardness of a distinguishing code can help in removing the error which in turn avoids attacks. Couvreur et al. [295] discussed the importance of indistinguishability under chosen-plaintext, chosen ciphertext, and adaptive chosen-ciphertext attacks in network coding-based post-quantum cryptography. Jäämeri [182] discussed the importance of a code-based cryptosystem with those schemes that are protected from structural weaknesses and Overbeck attacks. However, indistinguishability is a major challenge. The required security levels with certain overhead and randomness in the encryption scheme can be achieved. Singh [183] discussed McEliece, Niederreiter, and Classic McEliece-based cryptosystems. The cryptosystems based on these schemes are well protected from various attacks with different coding schemes. However, the hardness of the rank metric is required to ensure security in various applications like post-quantum cryptography security with network coding [293]. Bardet et al. [184] discussed the algebraic attack on the rank metric in a code-based cryptosystem. In this study, McEliece, and Niederreiter cryptosystems are used. It has been observed that weakness in rank metrics can result in algebraic attacks. Thus, there is a need to consider the hardness of a code in a rank metric. In an alternative solution, complexity bound could be applied to ensure the hardness of rank metric which in turn ensures the security of a cryptosystem.

**Table 9:** Analysis of contributions in Code-based cryptosystems

| Author | Cryptosystem | Error-Correcting Codes (ECC) | Major Strengths |
|---|---|---|---|
| Jäämeri [182] | McEliece, Gabidulin-Paramonov-Tretjakov (GPT) | Rank ECC, Gabidulin codes, Twisted Gabidulin codes, Twisted Reed-Solomon Codes | Protected from structural weaknesses and Overbeck's attack |
| Singh [183] | McEliece, Niederreiter, Classic McEliece, | Linear codes, Goppa codes | Strongly protected against brute force attacks.<br>Lesser disclosure of secret information can protect the schemes from total break, global deduction, local deduction, information deduction, and distinguishing algorithms. |
| Bardet et al. [184] | McEliece | Gabidulin codes, Reed-Solomon codes, Linear codes | Identified an attack that is below the security level for all rank-based schemes available in NIST Post-Quantum processes.<br>The proposed attack is useful for systems having small to medium scale parameters that require lesser memory compared to the best quantum attacks. |
| Ezerman et al. [185] | McEliece, Niederreiter | Goppa codes | A secure signature scheme is designed using a code-based cryptosystem. It is observed that the signature schemes in a code-based cryptosystem can be classified as "hash-and-sign" or "Fiat-Shamir".<br>The proposed scheme is a group signature scheme that requires multi-layered operations for generating group signatures. |
| Fernández-Caramés et al. [186] | McEliece, Niederreiter | Goppa codes, Low-Density Parity-Check (LDPC), Moderate-DPC (MDPC), Quasi Cycle Codes (QCC), Quasi-Cyclic Low-Rank Parity-Check (QC-LRPC), LRPC, LDPC | Conducted an in-depth survey of various post-quantum cryptosystem approaches and their variants.<br>The survey is focused on protecting the IoT systems using post-quantum computing. Further, IoT architectures and challenges are analyzed for providing guidelines to secure future post-quantum IoT systems. |

### 5.2.3 MULTIVARIATE CRYPTOSYSTEM

In multivariate cryptosystem, NP-hard and NP-complete multivariate equations are considered. The efficiency of a multivariate cryptosystem is based on the difficulty level in solving the systems of quadratic equations over a field. The concept of "One-way functions" composes of multiple easily invertible maps that could result in a difficult to invert function without much knowledge of individual sub-function in composition. Multivariate cryptosystem has many mature systems compared to other post-quantum cryptosystems because it started much earlier. The major advantage of multivariate cryptosystem includes fast processing, less computational and communicational resource requirements [188], and small signature generation in lesser polynomial time. Multivariate cryptosystems are largely classified into a digital signature, encryption/decryption, and other public-key cryptosystem-based approaches. In [189], NIST first round process is explained. There are a total of three rounds so far in post-quantum cryptography. In the third round, 7 finalists and 8 alternatives are selected for post-quantum cryptography. Cartor [190] discussed multivariate cryptography, the important direct algebraic attacks, differential techniques, and proposed a new multivariate encryption scheme. The proposed scheme is analyzed against algebraic, MinRank, discrete differential, and parameter selection-based attacks. The theoretical analysis gives a detailed picture of the multivariate scheme. However, practical aspects and their analysis is missing. Thus, this work can be extended to analyze the implementation aspects, performance analysis, and integration with application scenarios. Smith-Tone and Tone [191] studied the random linear code scheme-based nonlinear multivariate cryptosystem. This work has integrated the code-based and multivariate-based post-quantum cryptosystems. Thus, maximum security advantages can be taken out of it. Although this work has tested the proposed approached against various attacks this work can be extended to consider testing against weaknesses of code-based cryptosystems like hardness in indistinguishability of a code in a rank metric. In [195], the integration of a multivariate scheme with Blockchain is proposed. Here, an elliptic curve-based digital signature scheme and the Rainbow algorithm are used for creating a Blockchain. Security levels are varied from 80 bits to 256 bits and signature size, public and private key size variations are observed for two algorithms (Rainbow and Elliptic curve based digital signature scheme). This work has proposed the Ethereum network for analysis. However, this work can be extended to explore the private, public, and consortium-based Blockchain network for specific applications. Table 10 shows an analysis of multivariate cryptosystems.

**Table 10:** Analysis of Multivariate cryptosystems

| Cryptosystem | Variants | Major Strengths |
|---|---|---|
| Multivariate digital-signature schemes [189]-[192] | Rainbow digital signature schemes, Tame Transformation Signature (TTS), Tractable Rational Map Signature (TRMS), GeMSS, LUOV, MQDSS, Oil and Vinegar, Unbalanced Oil and Vinegar, Rainbow, CyclicRainbow, RainbowLRS2, Circulant Rainbow, NC-Rainbow. | Multivariate digital-signature schemes are comparatively more secure than multivariate encryption/decryption or public-key cryptosystem because short signatures are difficult to solve in polynomial time.<br>Simple arithmetic operations (addition and multiplication) make the schemes much efficient, especially for low-cost devices.<br>Multivariate schemes are considered to have very high security with small signature length. For example, the GeMSS scheme is found to achieve the NIST PQCSP level V security standard in the first round. |
| Multivariate Encryption/decryption schemes [190]-[193] | EFLASH, C* Toy, PFLASH, C*, SFLASH, Hidden Field Equation (HFE), $HFE^-$, ABC, SRP, EFC. | Multivariate encryption/decryption schemes are considered to be secure if they are protected against differential techniques, MinRank and algebraic attacks. |
| Multivariate public-key cryptosystem [194][195] | Multivariate Public Key Cryptosystem, Rainbow Signature Scheme | In this system, public keys are a set of polynomial defined over a finite field. Infinite field, the degree of the polynomial is often considered as 2. Thus, it is referred to as multivariate-quadratic cryptography as well.<br>Most of the multivariate public-key cryptosystems are quantum-resistant because no quantum algorithm solves the multivariate quadratic problem in polynomial time. |

### 5.2.4 ISOGENIES ON SUPER-SINGULAR-BASED CRYPTOSYSTEM

Cryptosystem-based on super-singular isogenies is an active area of research in post-quantum cryptography. The security of all supersingular isogeny cryptosystem schemes depends on the difficulty of computing the endomorphism ring of supersingular structures. Three popular isogeny-based structures used in post-quantum cryptography include Ordinary Isogeny Diffie-Hellman (OIDH), Supersingular Isogeny DH (SIDH), and Commutative SIDH (CSIDH) [196]. Using these structures, the protocols in isogenies on the super-singular cryptosystems are majorly classified as signature/encryption, key exchange, and hash function. Isogeny-based digital signature schemes ensure message integrity, nonrepudiation, and identity authentication. The core idea of ensuring these cryptography properties is to transform identification schemes into signature schemes with non-interactive zero-knowledge proofs. The challenges in the signature can be generated using hash functions. In key exchange protocols, public and private keys are used to generate session key that ensures confidentiality and integrity of subsequent communications. The hash function ensures collision resistance and compression. The challenges that need to address in the future include the use of new quasi-linear algorithms for isogeny evaluations, optimization in the finite field arithmetics for isogenies, avoiding inversions using projective curve equations, and use other optimization approaches (like Montgomery forms). In attacks, isogenies on super-singular-based cryptosystems should consider the design of those cryptosystems that are well protected from ephemeral key recovery, active attacks (like protecting the long-term keys), and side-channel attacks. In [197], the problem of endomorphism ring computation for supersingular elliptic curves is studied. This study is extended with an analysis of collision attacks over hash function parameters. The proposed directions to handle issues are generic in nature and can be extended for applications applying supersingular isogeny graphs. Thus, addressing Deuring's correspondence from maximal orders, or supersingular invariants can handle the preimage and collisions issues associated with a hash function or related parameters. In another scenario [198], the possibilities of power active attacks because of limited computing capabilities for the endomorphism ring of a supersingular elliptic curve are studied. In another major contribution, the factor involving partial knowledge in generating shared keys to determine the entire key is studied. This analysis is important to study side-channel attacks. Here, all forms of contributions are linked with computing capabilities. A higher computing capability and partial knowledge of keys can exploit the supersingular isogeny curves. In [199], an efficient commutative supersingular isogeny-based Fiat-Shamir signature algorithm is proposed. In this work, the large size of the public key is addressed by reducing it to half without affecting the security of the scheme. The proposed approach is tested against the quantum random oracle model and it is found to be secure for this scheme. Additionally, the proposed approach is found to be secure and effective compared to the existing approach in signing and verification. In verification, the challenge lies when there is a combination of the ephemeral key, secret key, and computational challenge. Thus, there is a need to address this challenge in the proposed scheme with fast, efficient, and security matters in consideration. In [200], another quantum adversary resistant signature scheme is proposed and it is named as 'Undeniable Blind Signature Scheme (UBSS)'. Although it has been analyzed that the proposed scheme is hard to solve, it does not address the issue of combination. Addressing a combination of keys and challenges with a blind signature scheme is important to take up. In [201], another blind signature scheme has been proposed. This scheme handles the undeniable signature issue in blind signature schemes. The proposed scheme is tested and found to be secure against various challenges. However, the issue of the combination of keys and challenges in the multi-party system needs to be taken up for further analysis. In [201[202], the importance of hash function, challenges, and efficient approaches are proposed and discussed. For example, Doliskani et al. [201] proposed a faster cryptographic hash function from supersingular isogeny graphs. The proposed approach provides exponential speed proportionate to characteristics of a finite field. The proposed approaches are claimed to be secure and less complex. However, an analysis against various active and side channel attacks can be conducted to work this work and ensure the security levels. Further, standard assumptions against whom the proposed approach is claimed to be secure should be used in comparative analysis with other similar work. The protocols in isogenies on the super-singular cryptosystem are briefly analyzed as shown in Table 11.

Table 11: Analysis of isogenies on the super-singular cryptosystem

| Category | Variants | Major Strengths |
|---|---|---|
| Isogeny-based signature/encryption algorithm [197][198]-[201] | SeaSign, CSI-FiSh, Quantum-resistant undeniable blind signature scheme, isogeny-based designated verifier blind signature scheme. | Lack of practices in the isogeny-based signature scheme makes this category of protocols weaker in post-quantum cryptography. |
| Isogeny-based key exchange protocol [197] | Longa, LeGrow, Galbraith, Authenticated Key Exchange (AKE)-SIDH-2, AKE-SIDH-3, SIDH-UM, biclique-SIDH. | The major challenge in key exchange protocol is to design authenticated key exchange protocol and verify the security with well-known security models such as BR, CK $CK^+$ |
| Isogeny-based Hash function [202][203] | CGL, Very Smooth Hash (VSH), VSH-DL, SWIFFT, Takashima's hash function, Charles, Goren and Lauter's hash function, | High-speed isogeny-based Hash functions are protected from Pollard-rho, claw finding, preimage, and collision-resistant attacks.<br>High-speed short messages-based Hash functions are useful to avoid quantum attacks of computational overhead that are used with novel solutions. |

### 5.2.5 HYBRID SCHEMES

In hybrid schemes, different post-quantum cryptography primitives and protocols are integrated to achieve set goals. For example, Crockett et al. [204] proposed a hybrid key exchange and authentication mechanism in Transport Layer Security (TLS) and Secure Shell Hash (SSH) protocols. The adoption of post-quantum cryptography with these mechanisms is found to be dependent on the standard of communication and availability of infrastructure. The integration of post-quantum and hybrid key exchange and authentication lies over the negotiation of multiple algorithms in hybrid cryptography that combine multiple keys and other primitives and protocols. The hybrid approach is found to be possible with the different hybrid key exchanges such as TLS 1.2. TLS 1.3 and SSHv2. Campagna et al. [205] proposed the integration of independent key exchanges and feeding mechanisms with Pseudorandom Function (PRF) to drive a secret and secure exchange. In this work, a new hybrid key exchange mechanism is designed for TLS 1.2 protocol with elliptic curve Diffie-hellman protocol and post-quantum key encapsulation. Further, Bit Flipping Key Exchange (BFKE) and Supersingular Isogeny Key Exchange (SIKE) are combined with the key exchange in TLS 1.2 handshake mechanism. Overall, the integration is found to be effective, and desired goals are achievable with food performance measures. Qassim et al. [118] combined physical layer and cryptography security primitives for increasing the security standard and proposed a cross-layer key agreement scheme that is strongly protected against a man-in-the-middle attack. This technique is found to be unbreakable and scalable to traditional cryptography primitives and protocols.

## 6. SCALABLE QUANTUM COMPUTER HARDWARE

As a full-fledged field, experimental quantum computing started as early as the 1980s, however, until the late 1990s, the majority of the researcher's envisaged industrial quantum computer as a distant reality [3]. Several contenders have attempted to create building blocks of a scalable quantum computer and they are developed independently by different academic researchers and industry engineers worldwide. For the design and implementation of qubits and quantum gates, a number of candidate material systems are being investigated. Some of the front-runner material systems include trapped ions [10], optical lattices [12], solid-state spins [11], electron spins in gated quantum dots [206], quantum wells [207], quantum wire [208], Nuclear Magnetic Resonance (NMR) [209], solid-state NMR [210], molecular magnet [211], cavity quantum electrodynamics [212], linear optics [213], diamond [214], Bose-Einstein condensate [215], Rare-earth-metal-ion-doped inorganic crystal [216] and Metallic-like carbon nanospheres [217] amongst others. However, superconducting circuits have transpired as the most widely used and successful material system to-date, although trapped ion system is also demonstrating excellent qubit fidelities and gate times.

The two main approaches for the physical implementation of a quantum computer are analog and digital [218]. A significant challenge for the construction of error-free industrial quantum computers is the maintenance of qubit state due to decoherence. Even with error rates achieved below 1%, the depth of quantum circuits required to solve real-world problems would be considerable, leading to detrimental cumulative error rates. Therefore, the area of quantum error correction is at present one of the most active areas of the research. Google Quantum AI, in collaboration with NASA, reported a demonstration of quantum calculation which was shown to require several thousand years on any conventional classical computer on 23 October 2019. Although this work achieved an important milestone for the current generation of quantum computers, the solution of a practical real-world problem on a quantum computer is expected to require significant further development. Notably, the work from IBM researchers showed that the efficiency of the same calculation on a classical supercomputer can be significantly improved [219].

### 6.1 QUANTUM COMPUTERS AND SPEED-UP

Quantum computers can solve the certain computationally intense tasks in significantly less time compared to classical computers, which is shown by the demonstrated "quantum supremacy". Another important term commonly used in the quantum community is "quantum advantage". While "quantum supremacy" implies solving a problem on a quantum computer which is intractable on any classical machine; whereas "quantum advantage" is a more practical term which deals with solving a useful real-world problem which cannot be efficiently solved on a classical computer. Although quantum supremacy has already been demonstrated, it is yet an open area of research to find practical problems which can be efficiently solved on quantum computers.

The quantum machines that have been engineered hitherto are bulky and offers limited computational power as they are made up of materials which have to be kept at superconducting temperatures, nevertheless, the potential of industrial quantum computers in future cannot be contested [218]. The motivation for potential benefits of industrial quantum computers can be derived from the present-day success of classical computers and the way they took off in the 1950s. Similar to the practical state of quantum computers today, the first generation of classical computers used to be bulky and had to be cooled continuously. As the theory of Artificial Intelligence (AI) had started shaping from the early days of classical computers, albeit they were nowhere near the compute required for AI, powerful industrial quantum computers can be theorized to come to reality in near-future and achieve "quantum advantage".

## 6.2 INDUSTRIAL APPLICATIONS OF QUANTUM COMPUTERS

Cryptanalysis is an inquiry into the information systems to determine the secret aspects of the system. It is used to circumvent the cryptographic safety mechanisms to access the contents of encrypted messages. An example is the RSA (Rivest–Shamir–Adleman) encryption which is widely used for encrypting data communication with banks and other nodes on the internet. Shor developed a quantum algorithm in 1994 which can, in principle break the operational RSA encryption if a large-scale error-corrected quantum computer can be developed. Hence, post-quantum encryption methods need to be formulated which can withstand an industrial quantum computer [43]. Searching efficiently and sorting through large data sets is now a high priority for many big enterprises. Grover developed an optimal quantum algorithm in 1996, which can speed up search through big data relative to the classical algorithms in query complexity. The present-day database software's such as Oracle are not suitable enough for real-world search enough to run Grover's algorithm; hence software that does the work of oracle in the quantum world need to be developed [44].

A variety of areas in computational sciences such as numerical weather prediction, computational chemistry and others involve solving equations using approximate methods ignoring the fine details. An example is the parameterization techniques used to approximate the sub-grid scale processes in a weather/climate prediction model due to the computational constraints. These approximate parameterizations have been known to propagate errors in the solutions to the system of equations, thus directly affecting the decision making. Industrial quantum computers offer hope in solving the equations in their exact form. This could for example allow an understanding of how different chemicals make fertilizers and improve upon the current high carbon footprint technique of manufacturing. Understanding chemistry, photosynthesis, superconductivity and magnetism, all being quantum mechanical phenomena can be better understood by industrial quantum computers. Although a scalable industrial quantum computer has still not been achieved and may require significant further development, research at the proof-of-concept level has started using the available, relatively less powerful quantum computers. On a seven-qubit quantum processor, IBM recently simulated beryllium hydride molecule [220]. Various applications such as patient diagnosis by quickly comparing the reports with a global database, modelling of live passenger and commercial traffic, the balance of energy supply and demand are expected to gain traction in the next few years. On the other hand, several other areas such as encryption, communications, financial transactions, critical infrastructure, Blockchain and cryptocurrency are some of the applications which are bound to become vulnerable by the development of an industrial quantum computer.

## 6.3 HARDWARE REQUIREMENTS OF INDUSTRIAL QUANTUM COMPUTERS

International efforts on how to build, construct and monitor qubit systems by over 100 academic and government-affiliated labs are underway. A number of large corporations and numerous ambitious start-up companies are now working on manufacturing of industrial quantum computers. Beside the development of qubits and quantum gates, an industrial quantum computer would also require intricate classical control and circuitry such as the application of electromagnetic fields, cooling system, user interface, networks and data storage capabilities. The hardware requirements of industrial quantum computers can be divided into four layers based upon their functions, viz the "quantum data plane", the "control and measurement plane", the "control processor plane" and the "host processor". The "quantum data plane" is the location where qubit states are stored and measurements are carried out by the "control and measurement plane". The sequence of operations in algorithms are taken care of by the "control processor plane", and the "host processor" carries out the user interface, networks and storage of large arrays.

## 6.4 CHALLENGES IN SCALABLE PRODUCTION OF INDUSTRIAL QUANTUM COMPUTERS

In order to build a functional industrial grade quantum computer, several technological issues have to be addressed; the most important of which being the detrimental impact of noise or decoherence which causes errors in quantum computation and suppresses quantum advantage. An initial state of a qubit has to be set before it can be used in addition to developing circuits and gates. Photons remain coherent for a long time; however, creating quantum circuits out of them is a challenge. Superconductors possess quantum properties which can be harnessed to develop quantum circuits which are in use by IBM, Google, Rigetti and others to build their quantum computers. However, the fidelity of these qubits, in particular of two-qubit operations is still relatively low and therefore require error correction or mitigation techniques to be implemented. In 2016, IBM released a five-qubit processor free for everyone on the cloud, which can be used to construct a quantum circuit and run it as long as it uses five or fewer qubits [221]. At present, IBM offers cloud access to quantum computers consisting of up to 65 qubits and have recently announced a quantum computer with a record 64 quantum volume [222].

Table 12 shows five major candidate material systems for the development of an industrial quantum computer and the relevant metrics to measure their performance and the current state-of-the-art. Among these candidate systems, Trapped Ion and Superconducting qubits are the basis for the current generation of quantum machines available through cloud access. The other three material systems are still a subject of intense research and require significant further development to be available for quantum circuit simulations [223]. Although there has been much progress in designing smaller quantum computers, it is not yet possible to experimentally demonstrate a design for an industrial quantum computer which could be of the scale required to crack current cryptography and the existing implementations even if scaled up are not just enough. Scaling the qubits to achieve an industrial quantum computer has many challenges such as the quality of qubits when scaling up to industrial-scale quantum computers, wiring, refrigeration, packaging and others.

Table 12: Major hardware candidates for industrial quantum computer and their properties.

| **Qubit Technologies** | **Trapped Ion Qubits [299]** | **Superconducting Qubits [300]** | **Silicon Qubits [301]** | **Photonic Qubits [302]** | **Topological Qubits [303]** |
|---|---|---|---|---|---|
| Physical Qubits | IonQ:79; AQT:20 | IBM: 65 qubits; Google: 54 qubits; Rigetti: 30 | 2 | $6x3^9$ | In progress |
| Coherence Times | ~50 sec | ~50-200 μsec | ~1-10 sec | ~150 μsec | - |
| Gate Fidelity | ~99.9% | ~99.4% | ~90% | ~98% | Expected: ~99.9999% |
| Gate Operation Time | ~3-50 μsec | ~10-50 nsec | ~1 nsec | ~1-10 nsec | - |
| Scalability | Some potential | Medium to high potential | High potential | High potential | - |

Theoretically, silicon-based quantum computers have been predicted to offer the potential for scalability with error correction schemes. After the seminal work from Kane in 1998 [224], many surface-code quantum computer architectures have been proposed [225] [226] [227]. Remarkable advancements in silicon spin qubit design and characterization [228] [229] [230] [231] [232] [233] demonstrated in the recent years, confirm the suitability of this material system as an attractive candidate for the construction of a scalable industrial quantum computer.

## 6.5 CURRENTLY AVAILABLE PLATFORMS

IBM released the quantum computer known as IBM Quantum Experience in 2016 which was a five-qubit system. The system was launched with a user guide and a community forum. Later in 2017, a number of features were added to IBM Quantum Experience such as giving permissions to the users to interact via quantum assembly language, interactive use interface and simulator expansion. IBM then launched Qiskit which helped to code on the quantum processor. Further they developed a 16-qubit system and also launched the quantum awards program. The Quantum Experience is a cloud-computing based platform which provides access to the public to the quantum processors, an online forum and the tutorials to code on Q devices of IBM. Various research publications have used the IBM Quantum Experience. The hardware of quantum processors by IM is superconducting qubits which reside inside a dilution refrigerator. The Graphical User Interface (GUI) that users interact with is known as the quantum composer. Quantum composer is used to write quantum assembly code. The GUI facilitates the development of quantum experiments and algorithms. The option to use a real processor or a simulator is also available.

A similar cloud-based quantum computing service is provided by Rigetti Computing through its platform known as Forest. The company is primarily known for manufacturing quantum integrated circuits. Forest helps the coders to access the cloud-based quantum processors by Rigetti wherein they can test their quantum algorithms. They have also developed a dedicated quantum instruction language called Quil which is used for the cloud-based quantum computing as a service. More than 36 qubits are available on a quantum chip of Forest and Python programming can be used for hybrid classical or quantum computing. Quantum Inspire is a Europe based cloud-computing based quantum platform which is providing its services under the name of a company known as QuTech. The cloud-based quantum computing systems offer access to the power of quantum computing and simulate quantum algorithms without the need of buying or building a quantum computer.

### 6.6 STATE-OF-THE-ART AND FUTURE OUTLOOK IN INDUSTRIAL QUANTUM COMPUTERS

The size of the industrial quantum computing market is expected to touch $ 1.9 bn by 2023 and $ 8 bn by 2027 [234]. Various computing giants such as IBM, Microsoft, Alibaba, and Google dedicated quantum enterprises such as D-Wave and others such as Rigetti Computing and NVision Imaging Technologies are testing quantum computers competing to launch the scalable industrial computer. Global research and development efforts are ongoing to commercialize industrial quantum computers with continuously increasingly leading contributions from US and other prominent efforts coming from the EU quantum technologies flagship and the UK national quantum technologies program, the Australian Centre for Quantum Computation and Communication Technology (CQC2T) and the Chinese quantum national laboratory for quantum information science.

### 6.7 BLIND QUANTUM COMPUTATION

Blind Quantum Computation (BQC) ensures infrastructure to do quantum computations while hiding from the server the computational structure. In addition to privacy protection, many BQC techniques incorporate embedded control tests that check the computation process. BQC allows us to do calculations without revealing the calculation results to anybody. In BQC, the encryption protocol safeguards computational inputs, outputs, and algorithms [307] [308] [309]. Homomorphic encryption encrypts inputs and outputs only. Thus, BQC is more secure than homomorphic encryption. Figure 12 shows the important terminologies associated with BQC. Some of the important concepts related to BQC are briefly explained as follows:

- Universal Blind Quantum Computation: UBQC protocols permit the client to create random states from a finite set. These states are used to ensure secure delegation of quantum computational tasks to a server. UBQC protocol consists of four phases, including pre-computation (for angles measurement and unitary computation from brickwork state), Alice's preparation (qubits computation), Bob's preparation (brickwork state computation), and interaction and measurement (angle measurement, angle encoding, and verification process). The performance of protocol is measured using correctness, universality, and security.
- Blind Oracular Quantum Computation (BOQC): In BOQC, a third party executes the client's oracular quantum computations on a server. Here, third-party support is considered because the client is assumed to have limited quantum power and capacity to construct an oracle. Third-party identifies a server that can take the help of an oracle to do the required computations. Gustiani et al. [310] surveyed important concepts, protocols, and terminologies associated with BOQC in the quantum era.
- BQC Protocols and Parameters: In addition to BQC, UBQC, and BOQC, various protocols and parameters are associated with BQC. For example, single-server BQC protocols, double-server protocol, triple-server protocol etc. Figure 12 shows the classification of important protocols and parameters in the BQC area [307][308][309].

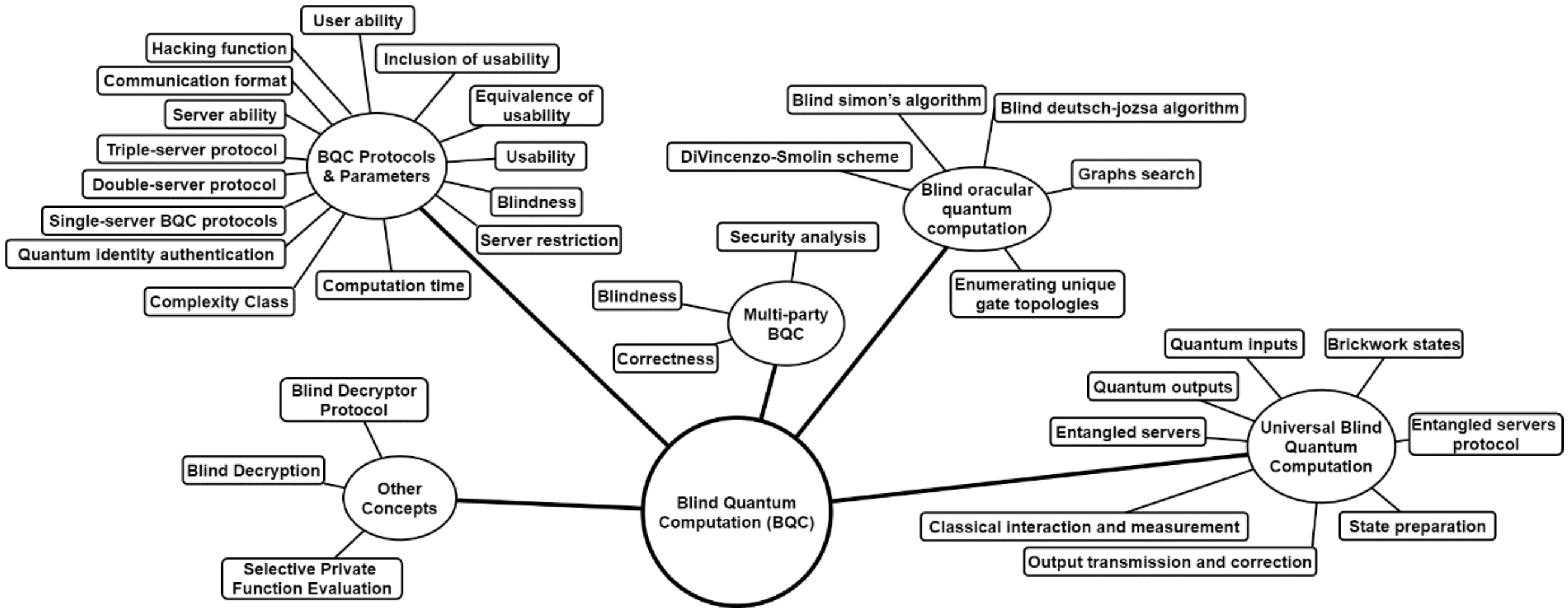

**Figure 12:** Blind Quantum Computation and Taxonomy

## 7. FUTURE DIRECTIONS

We have identified various ongoing research areas in quantum computing for three different maturity levels (5 years, 5 to 10 years and more than ten years) based on the state-of-the-are research. They are illustrated in Figure 8 as the hype cycle for quantum computing. *T* represent Technology, and *A* means the Application area in the Hype Cycle. As per Figure 13, post-quantum cryptography is at the peak while a lot of research work has been done on simulations for complex quantum experiments. Research areas such as Robotics, energy management, cybersecurity, distributed QC, complex computational chemistry, financial modelling, and drug design are at the kickoff stage in their development under the domain of quantum computing. The use of quantum computing in these areas at their innovation trigger, may take more than 10 years to mature. Traffic Optimization is also at its innovation trigger but is expected to top the hype cycle within the next 5-10 years. Quantum cryptography, quantum control, and adiabatic quantum computing have peaked inflated expectations. It is expected that it would take less than five years for them under complete development under quantum computing purview. Quantum Internet, quantum-based satellite communication, quantum assisted machine learning, electronics material discovery, and error-corrected quantum computing have also reached the peak of inflated expectations but are anticipated to rapidly evolve in five to ten years. Quantum based portfolio-risk optimization and fraud detection, and fault-tolerant quantum computing presently have high expectations on the hype cycle and are blossom in more than ten years. A lot of research work has been done on quantum algorithms & complexity and quantum programming languages & systems, which could be active research areas during the next 5 to 10 years. Simulation software for quantum experiments and quantum simulators is at the enlightenment slope and has a long way to develop in quantum computing fully. We have identified various open challenges and future research directions, which are still a topic of active global research.

### 7.1 ENGINEERING/DESIGN CHALLENGES

Fragility is the main drawback of Quantum technology due to two following reasons [235]: 1) a short coherence time of qubits because superconducting qubits forget their information very frequently (in nanoseconds). 2) There is unreliability in quantum operations due to relatively large error rates, and it is challenging to develop a quantum computer with low error rates. Moreover, small material faults or environmental instabilities can generate an error in qubits and lose their quantum data, which reduces the useful period of a qubit. There is also need to perform logical functions while controlling the qubit to reduce incidental electromagnetic noise, which can decrease decoherence. To improve the scalability of the quantum computers, balance is required among protecting qubits from prospective environmental instabilities. Further, as compared to classical computing, error correction in quantum computing is quite challenging because a) errors are continuous (involve both amplitude and phase), b) cannot copy unidentified quantum states and c) measurement can collapse a quantum state and destroy the data saved in Qubits. To run a quantum algorithm efficiently, many physical qubits are required, which need a close and continuous connection between the classical platform and quantum chip, and it forms a colossal control overhead [236]. Moreover, this interaction and overhead increases the complexity for quantum computing process in terms of run-time control, architecture and integration. Currently, qubit count is using to measure the power in quantum computing hardware. Still, this measure is not giving correct value and leads to the challenge of the power of future powerful quantum computers with more than 1000 qubits. Qubit architecture improves scaling to solve the dynamic sized complex problems, but it needs an efficient cooling component to maintain heat, which can be solved by utilizing AI-empowered systems.

### 7.2 RELIABLE QUANTUM COMPUTING

It is challenging to attain fault-tolerant and reliable quantum computations as practical implementation of quantum error correction is still an open problem [237]. Due to quantum states' delicate nature, there is a need to operate bits at very low temperatures, and fabrication should be highly accurate [238]. It is also challenging to measure complete quantum state accurately; therefore, verification is challenging. There is a significant probability of errors during computation as compared to classical computing. There is a need for an effective error correction mechanism for quantum architectures to operate as intended. There is also a need to redesign quantum communication architecture to increase the verification of precise fabrication constraints. On the other hand, qubits are very difficult to test after fabrication because tolerances are tight, and the use of incorrectly placed Qubits must be avoided to reduce the occurrence of error. There is a need to apply error correction recursively to attain adequate fault tolerance to permit sustainable quantum computation [239]. In future, the latest AI and ML-based techniques can be used for automatic detection and corrections of errors dynamically to offer valuable and reliable service. The utilization of recent AI and ML-based techniques can improve the reliability but it can also increase the complexity within the system by increasing the processing of data, which leads to extra training cost for AI/ML techniques as well.

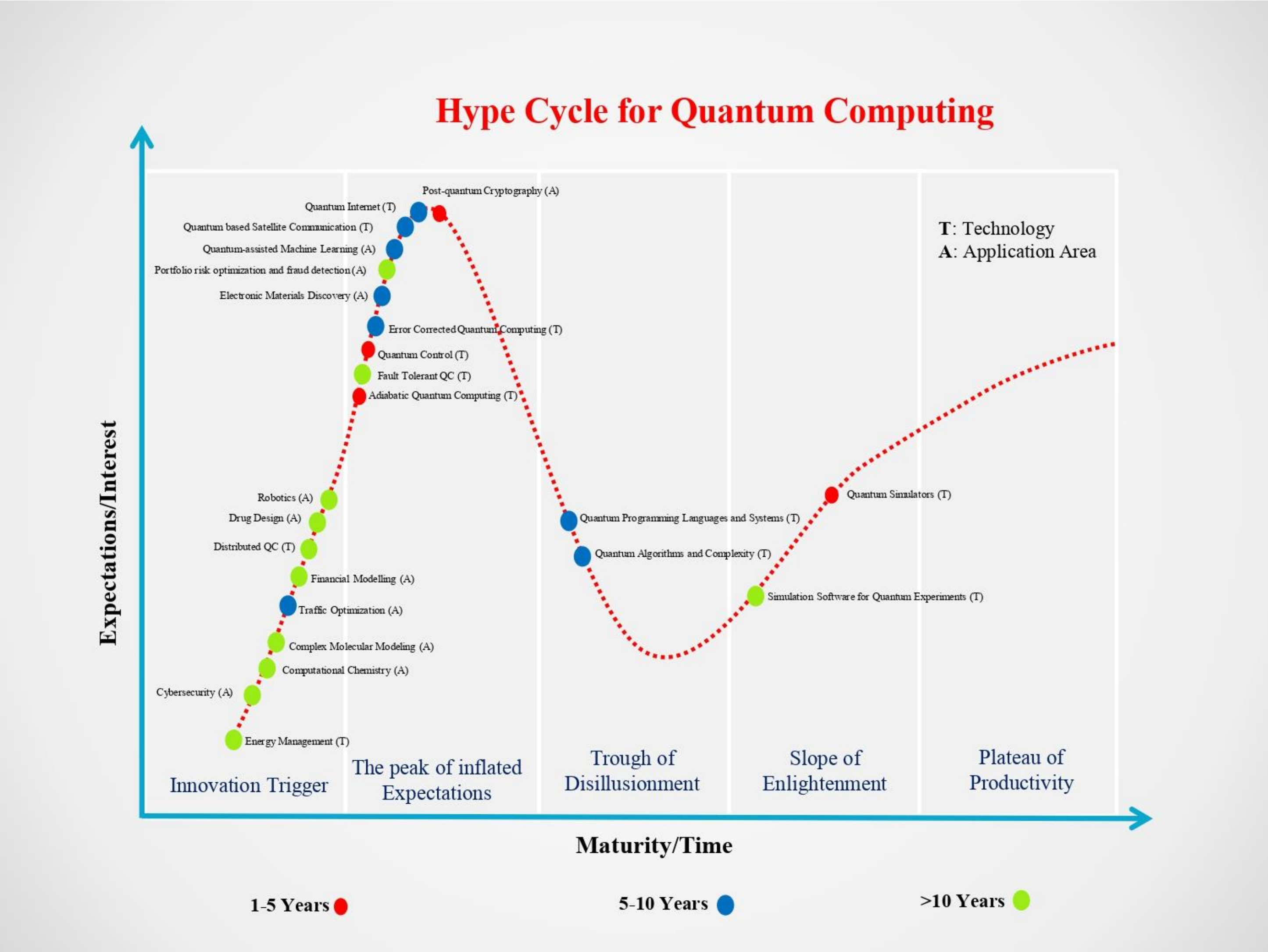

Figure 13: Hype Cycle for Quantum Computing

### 7.3 QUANTUM-ASSISTED MACHINE LEARNING

Machine learning researchers use principal component analysis, vector quantization, Gaussian models, regression and classification in routine [244]. To improve the scalability and efficiency of machine learning algorithms, quantum technology can be used in handling large datasets with large sizes of devices (100–1000 qubits) [241]. Further, quantum computers can efficiently attract the interest of the machine learning community by preparing and sampling definite probability distributions efficiently, such as training in classical and quantum generative models. Nowadays, the input size (the number of users) for the quantum recommendation system algorithms is increasing, which is challenging to complete the operation with the required speed. There is a need for millions of qubits to handle the current demand and tackle large datasets. The hybrid quantum-classical algorithms can solve this problem by providing current computation power and other machine learning tasks [241]. The other essential challenges can be limited qubit connectivity, and the device's integral noise increases decoherence in the qubits. The utilization of advanced AI and reinforcement learning can increase the scalability and offer more computational power to handle a vast amount of data generated from various IoT devices [240]. Further, NISQ devices with the viewpoint of tensor networks can be used to explore the workflow of quantum-assisted machine learning, which can provide a healthy platform to develop innovative ML models to improve the resource management within the quantum computer. The effective management of resources can also reduce the impact of the noise fluctuations on the performance of quantum hardware.

### 7.4 ENERGY MANAGEMENT

Energy management is a significant challenge, where the world's powerful supercomputers and Cloud Data Centres consume a lot of energy to solve different problems [242]. Quantum computers are expected to be more energy-efficient than them (supercomputers/Data Centres) while executing a particular task [243]. On the other hand, a quantum computer can reliably perform extensive calculations using less energy, which further reduces the cost and carbon emissions. Classical computers use binary bits (0 or 1) for encoding information, while quantum computing uses Qubits, which represent both 0s and 1s simultaneously—this property of quantum computing to identify an optimal solution while consuming less energy. The reason for less energy consumption is that the quantum processors are working at shallow temperature, and the processor is superconducting with no resistance, which means no production of heat [244]. Hybrid applications contain two portions: high-energy and low-energy [242]. Quantum computing executes the high-energy portion, while classical computing executes the low-energy part using the cloud [245]. To solve these kinds of problems, there is a need for hybrid computing comprised of quantum and classical computing to curb energy usage and costs dramatically. There is a need to do more work before implementing hybrid computing to solve today's most challenging business problems. Quantum computers can use AI to improve computational speed, reliability, and security and increase the size of infrastructure, which needs a vast amount of energy to run it and control the temperature using cooling devices. In future, the energy demand of these Quantum computers can be fulfilled with the utilization of renewable energy along with brown power. Further, the energy demand of quantum computers can be predicted using latest machine learning techniques to estimate the demand of both renewable and non-renewable energy. Further, effective data analytics techniques can be utilized to perform accurate predictive analytics for energy consumption. The quantum computers need to be scaled up from 50 qubit systems to the 10,000 for solving the complex problems of computational chemistry and biology, which needs more energy for computing and cooling (to maintain the temperature). So, there is a need to develop the energy-efficient quantum data centers for better utilization of energy.

### 7.5 QUANTUM INTERNET

Quantum Internet enables distributed quantum computing by incorporating new communications and improving computing capabilities to a large extent. Quantum Internet has various challenges because it uses quantum mechanics laws, and the main constraints for network design are teleporting, entanglement, quantum measurement and no-cloning [246]. The error-control mechanism is an essential assumption of classical computing, which is no more valid in quantum computing. There is a need for a major shift in network paradigm from classical to quantum specific to the design quantum Internet. Further, when a qubit interacts with an environment, it causes decoherence because Qubits are fragile and lose information from Qubit to the environment over time. Moreover, the long-distance entanglement distribution is also a challenge in quantum computing for the effective transformation of data. To improve the computation and communication mechanisms within quantum computer nowadays needs a large amount of memory which would be more challenging with future quantum Internet to retain the details of operations performed. Further, there is need of high bandwidth to offer effective communication among quantum devices, computers and web applications. It would be also challenging to make the current web applications compatible with quantum Internet applications depend on entangled qubits. So, there is a need of uniform interface which allow quantum sensors, quantum computers and quantum Internet applications can exchange data using quantum Internet.

### 7.6 ROBOTICS

Robots use GPUs to solve intensive computational tasks such as drug discovery, logistics, cryptography, and finance, where quantum computing can be augmented to perform computations with a considerable speed. Quantum-powered robots can also utilize cloud-based quantum computing services to solve different types of problems [247]. Nowadays, quantum computing enhances robotic senses for manufacturing, such as identifying several faults in a jet engine in a short period [248]. Further, Quantum image processing helps to understand the visual information efficiently and saves and manages image data effectively using two critical quantum computing properties such as parallelism and entanglement. Artificial intelligence-based robotics are dealing with different kinds of problems using graph search to deduct new information, but complexity increases with the increase in data. Quantum computing can reduce complexity by using quantum random walks instead of graph search. Further, other significant problems related to kinematics, such as the mechanical movement of robotics, can be solved by quantum neural networks by enhancing machines' activity and recognizing moments of joint friction and inertia. Moreover, another problem, such as identifying the reason for the inconsistency between expected and observed behaviour, is challenging, which could be solved using Quantum algorithms. Quantum computing uses to optimize the motion of machines in robotics, such as joint friction and moments of inertia, which can be solved by quantum reinforcement learning in the future. The utilization of quantum processor is improving the automatic learning process in robotics using superposition principle but it is also increasing complexities within the system. Further, it will also increase the cost of building quantum technology based robots due to training and learning expenses of machine learning models.

### 7.7 SIMULATIONS FOR COMPLEX QUANTUM EXPERIMENTS

Quantum computing can simulate complex chemistry, physics, and biology problems using small-scale (50-100 qubit) 'quantum simulators', which could be available in coming years [249]. The expertise of an extensive range of researchers and fundamental aspects of classical computing can work together to understand and harness quantum technology's capabilities. Further, quantum simulators can realize the natural system while solving complex problems (which is difficult to solve on the classical system or supercomputer) in a controlled manner to measure the influence of various parameters on each other. Quantum simulators can take advantage of quantum computing's essential properties such as entanglement and superposition while designing it. There is a need to develop large-scale programmable quantum system for effective processing of information for complex processes in Chemistry and Physics. In future, more scalable systems or simulators can be developed to run large-sized and complex jobs related to Chemistry and Biology with optimum results by investigating the hardware-efficient realization of quantum algorithms.

### 7.8 POST-QUANTUM CRYPTOGRAPHY

There is a need for cryptography to improve the security for implanted medical devices, cares, and online communication. Nevertheless, the various generally used cryptosystems will be damaged once large quantum computers come into existence. Post-quantum cryptography denotes the cryptographic algorithms (generally public-key algorithms). It is assumed that the attacker used a large quantum computer to attack in post-quantum cryptography, and these systems attempt to stay secure in this situation [46]. Post-quantum cryptography has to maintain integrity and confidentiality while preventing different kinds of attacks. Post-quantum cryptography research is typically concentrated on six techniques such as symmetric key quantum resistance, supersingular elliptic curve isogeny cryptography, code-based cryptography, hash-based cryptography, multivariate cryptography and lattice-based cryptography [250]. Another challenge within post-quantum cryptography is “Agility”; there is a need to find out the right areas to incorporate agility. Therefore, future systems should build in such a way, which must be able to predict the possible security problems. Further, there is a need for testing and validation design by developing new automated tools to identify and fix the fault at runtime dynamically. Moreover, the reconfiguration of legacy devices with cryptosystems is still an open problem, which needs to be solved by incorporating agility in the legacy applications. Future works need to explore more secure code-based systems that give outcomes at a lesser cost of delay. Thus, trade-offs between delay, security, and information rate need to be studied in detail. High computational and communicational rates without scarifying security are the aim. To adapt to post-quantum cryptography transition in real-time applications, there is a need to formalize a wide array of standards. For example, integration with banking, remote learning, mobile communications, healthcare, and other emergency services, and critical infrastructure requires studying post-quantum algorithm choices. The selection of these algorithms can speed up the migration process as well. Quantum computers can use Quantum Key Distribution (QKD) and integrates verifiable quantum key generation with quantum-safe cryptosystems (multivariate constructions, lattice-based, isogeny-based, hash-based and code-based) to provide the unbreakable security but it would be expensive. Another non-expensive solution could be hybrid implementations (pre-quantum and post-quantum schemes), which can also be used to improve the data privacy from quantum capable attacker.

### 7.9 NUMERICAL WEATHER PREDICTION

The development of classical computers was accompanied by advancements in numerical weather prediction skills in the 1950s. Since then, the predictions of weather forecasts have greatly enhanced in the last few decades. This development has been catapulted by the improved hardware and software but has been limited by the fundamental principle on which these traditional computers are built, i.e. bits or 0s and 1s. For the purpose of colossal calculations required, the classical computers are stacked to build what are known as supercomputers. These supercomputers perform computations day and night to generate forecasts of the atmosphere, ocean, land, and other components of the Earth system. Although they have improved with time, the state-of-the-art predictions still need a lot of upgrades for societal applications such as flood forecasting, urban modelling, sub-surface flow modelling, and allied complex tasks. These developments have been limited by the computational power available today. With the hope of industrial quantum computers becoming a reality, the next-generation Earth System Models would be able to run at much higher spatial and temporal resolutions. There is a need to diligently study quantum computing's applicability to numerical weather predictions [251]. Numerical weather prediction can adopt quantum computing because the limitations of classical computing lead to erroneous high-resolution forecasts. The scientific goal is partial differential equations on the three-dimensional spherical atmosphere and ocean, which is limited in the spatial resolution by the computational power of classical computers. Quantum computing can tackle the important challenges of climate change such as global warming and the production of $CO_2$ emissions. Further, weather and climate models can be simulated using quantum technology based large scale simulators to determine different catalysts for carbon capture in a cost-effective manner.

### 7.10 QUANTUM CLOUD COMPUTING AND CRYPTOGRAPHY

An unconditional secure quantum cloud computing can be a major ingredient to various real-life applications if powerful quantum computers will become widely available in future [252]. A few powerful quantum-computer nodes in a cloud would make the client's job much easier. Client would need to communicate with quantum clouds via a quantum link for transferring their job and associated qubits. The efforts have been made in this direction to experimentally demonstrate blind quantum computing where input, delegation, computations and output are unknown to quantum servers. These developments have been limited by the universal and powerful quantum clusters. Cryptographic verification of quantum cloud computing, fault-tolerant secure quantum computations, Error-free quantum cryptography mechanisms, cryptography primitives and key distribution mechanisms in quantum cloud computing environment, and quantum techniques for access control in cloud computing [253] [254] [255] [256]. In conclusion, secure and efficient quantum cloud computing environment is required to be studied in-depth for universal quantum computing at large scale. Further, cloud-based environment will be an effective approach for storage, computation and distribution of data to the quantum computing community. In these systems, latency and network bandwidth can be challenges for the execution of small jobs, which can be solved using the concept of fog/edge computing. To provide the quantum as a cloud service, there is a need of large scale systems which can offer autoscaling. Serverless computing can be used to offer the dynamic scalability to solve the complex problems along with quantum technology. The latest security mechanism such as Blockchain can be used to provide more secure and reliable service. Further, the integration of Blockchain service with Quantum Internet can improve the communication speed along with required security.

## 8. SUMMARY AND CONCLUSIONS

This paper presents a systematic review of quantum computing literature. It identified that quantum-mechanical phenomena such as entanglement and superposition are expected to play an important role while solving computational problems. We proposed a taxonomy of quantum computing and mapped it to various related studies to identify the research gaps. Various quantum software tools and technologies are discussed. Further, post-quantum cryptography and industrial quantum computers are discussed. Various open challenges are identified, and promising future directions are proposed. The fusion of all the performance attributes in a single quantum computing technique is still ambiguous until now. To build a quantum computer which can perform concurrent operations, it is essential to have a quantum computing technique that can allow quantum I/O with all the necessary classified features. The suggested taxonomies framework can be used to contrast various existing quantum computing techniques for determining the optimal strategy that can be applied on classical computing infrastructure. However, the scaling of qubits, trade-off between speed and the decoherence time is the topic of research in the field of quantum computing.

Quantum computers are developed to increase the security rate in communication and computations via decreasing the computational time. To secure the classical cryptography primitives and protocols with the usage of quantum computer's ability in solving the mathematical problems in few milliseconds, post-quantum cryptography mechanisms are designed. Post-quantum cryptography strengthens the symmetric cryptography primitives and protocols against well-known quantum attacks. Further, it has taken three hard mathematical problems (integer factorization, discrete logarithmic, and elliptic-curve discrete logarithm) in asymmetric key cryptography to secure the cryptography primitives and protocols. In conclusion, the characteristics of post-quantum cryptography increase the computational efficiency and security of many futuristic applications.

Furthermore, the present-day industrial quantum computers are not yet there to replace classical supercomputers owing to the challenges in scaling up on the number of qubits that can be practically realized hitherto. When that might happen is an open question. Though the next decade is going to be highly exciting for industrial quantum computers, there is still uncertainty on when the quantum computers will start to replace their classical counterparts in complex tasks. However, digital supercomputers are here to stay, even if quantum becomes a reality, as an addendum to the quantum computers of the future.

There is one crucial design challenge: how to run a quantum algorithm efficiently? A large number of physical qubits are required, which need a close and continuous connection between the classical platform and quantum chip, forming a huge control overhead. It is challenging to achieve fault-tolerant and reliable quantum computations because of quantum error correction, which is still an open problem. Due to the fragile nature of quantum states, there is a need to operate bits at very low temperatures and fabrication should be accurate. Further, to improve the scalability and efficiency of machine learning algorithms, quantum technology can be used in handling a large dataset with a large number of devices (100–1000 qubits). Energy management is a research area in the field of quantum computing. To improve energy efficiency, there is a need for hybrid computing comprises of quantum and classical computing to curb energy usage and costs dramatically. There is a need to do more work before implementing hybrid computing practically to solve today's hardest business problems. Quantum simulators can be designed for simulations for complex quantum experiments, which can take advantage of important properties of quantum computing such as entanglement and superposition while designing it. Presently, Artificial Intelligence based robotics are dealing with different kinds of problems using graph search to deduct new information, but complexity is increasing with the increase in data. Quantum computing can reduce the complexity of robotic mechanism by using quantum random walks instead of graph search. Other various fields such as computer security, biomedicine, the development of new materials and the economy will benefit from the advancement in quantum computing.

## DATA AVAILABILITY STATEMENT

Data sharing not applicable to this article as no datasets were generated or analyzed during the current study.

## ORCID

Sukhpal Singh Gill https://orcid.org/0000-0002-3913-0369

Adarsh Kumar https://orcid.org/0000-0003-2919-6302

Harvinder Singh http://www.orcid.org/0000-0002-9427-1279

Manmeet Singh https://orcid.org/0000-0002-3374-7149

Kamalpreet Kaur https://orcid.org/0000-0001-7162-2059

Muhammad Usman https://orcid.org/0000-0003-3476-2348

Rajkumar Buyya https://orcid.org/0000-0001-9754-6496

## Appendix A: List of Abbreviations

| Notation | Description |
|---|---|
| ABC | Simple Matrix Scheme or ABC in short |
| AKCN | Asymmetric Key Consensus with Noise |
| BB84 | Quantum key distribution scheme |
| Bi-GISIS | Bilateral Generalization Inhomogeneous Short Integer Solution |
| CHP | CNOT-Hadamard-Phase - Scott Aaronson |
| CK+ | Extended Cohn-Kanade (CK+) database |
| Cirq | Software library for writing, manipulating, and optimizing quantum circuits |
| DWDM | Dense Wavelength Division Multiplexing |
| EFC | Extension Field Cancellation |
| EMBLEM | Error-blocked Multi-Bit LWE-based Encapsulation |
| EQCS | Egyptian Quantum Computing Society |
| GeMSS | Great Multivariate Short Signature |
| HFE | Hidden Field Equation |
| ILWE | Integer Module Learning With Errors |
| KCL | Key Consensus from Lattice |
| LanQ | A quantum imperative programming language |
| LDPC | Low-density parity-check |
| LUOV | UOV + PRNG + Field Lifting + Simplified Secret Key |
| MDPC | Moderate Density Parity Check |
| MLWE | Module Learning With Errors |
| MPKC | Multivariate Public Key Cryptosystems |
| MPLWE | Middle Product Learning With Errors |
| MQDSS | MQ (multivariate quadratic) Digital Signature Scheme |
| NC-Rainbow | Non-Commutative Rainbow |
| OKC | Optimal Key Consensus |
| OKCN | Optimally-Balanced Key Consensus with Noise |
| PRNG | Pseudorandom Number Generator |
| R. EMBLEM | Ring Error-blocked Multi-Bit LWE-based Encapsulation |
| RLWE | Ring Learning With Errors |
| RDSS | Rainbow Digital Signature Schemes |
| R-LRS2 | Rainbow Low Resolution Spectrograph 2 |
| SRP | MPKC encryption scheme called SRP |
| SIS | Short Integer Solution |
| staq | Full-stack quantum processing toolkit written in standard C++ |
| TRMS | Tractable Rational Map Signature |
| TTS | Tame Transformation Signature |
| QCAD | Quantum Computer Aided Design |
| QPU | Quantum Processing Unit |
| QCGPU | Quantum Computing GPU |
| qchas | A library for implementing Quantum Algorithms |
| QC-LDPC | Quasi-Cyclic Low-Density Parity Codes |
| QC-LRPC | Quasi-Cyclic Low-Rank Parity-Check |
| QIO | Quantum Input Output |
| QMDD | Quantum Multiple-valued Decision Diagram |
| QOCS | Qualified One-way Costs Shifting |
| QuEST | Quantum Exact Simulation Toolkit |
| UOV | Unbalanced Oil and Vinegar |